\documentclass{aa}

\usepackage{color}
\usepackage{graphicx,natbib}
\usepackage{deluxetable}

\bibpunct{(}{)}{;}{a}{}{,}

\def\ergscm{erg~s$^{-1}$~cm$^{-2}$}

\def\arcmin{\hbox{$^\prime$}}

\def\flux{erg s$^{-1}$ cm$^{-2}$}

\def\aj{AJ}
\def\apj{ApJ}
\def\apss{Astroph.Sp.Sci.}
\def\aap{A\&A}
\def\mnras{MNRAS}

\newcommand{\comment}[1]{}

\newcommand{\lesssim}{\mathrel{\hbox{\rlap{\hbox{\lower4pt\hbox{$\sim$}}}\hbox{$<$}}}}
\newcommand{\gtrsim}{\mathrel{\hbox{\rlap{\hbox{\lower4pt\hbox{$\sim$}}}\hbox{$>$}}}}
\newcommand{\beq}{\begin{equation}}
\newcommand{\eeq}{\end{equation}}
\newcommand{\beqa}{\begin{eqnarray}}
\newcommand{\eeqa}{\end{eqnarray}}

\begin{document}

\title{INTEGRAL/IBIS all-sky survey in hard X-rays\thanks{Based on observations with INTEGRAL, an ESA project with
instruments and science data centre funded by ESA member states
(especially the PI countries: Denmark, France, Germany, Italy,
Switzerland, Spain), Czech Republic and Poland, and with the
participation of Russia and the USA} }

\author{R. Krivonos\inst{1,2}, M. Revnivtsev\inst{1,2}, A. Lutovinov\inst{2,1}, S. Sazonov\inst{1,2}, E. Churazov \inst{1,2}, R. Sunyaev\inst{1,2}}
\institute{
              Max-Planck-Institute f\"ur Astrophysik,
              Karl-Schwarzschild-Str. 1, D-85740 Garching bei
              M\"unchen, Germany
\and
              Space Research Institute, Russian Academy of Sciences,
              Profsoyuznaya 84/32, 117997 Moscow, Russia
            }

\authorrunning{Krivonos et al.}

\abstract{

We present results of an all-sky hard X-ray survey based on almost
four years of observations with the IBIS telescope on board the
INTEGRAL observatory. The dead time-corrected exposure of the survey
is $\sim33$~Ms. Approximately 12\% and 80\% of the sky have been
covered to limiting fluxes lower than $1$ and $5$~mCrab, respectively.
Our catalog of detected sources includes 400 objects, 339 of which
exceed a $5\sigma$ detection threshold on the time-averaged map of the
sky and the rest were detected in various subsamples of
exposures. Among the identified sources, 213 are Galactic (87 low-mass
X-ray binaries, 74 high-mass X-ray binaries, 21 cataclysmic variables,
6 coronally active stars, and other types) and 136 are extragalactic,
including 131 active galactic nuclei (AGNs) and 3 clusters of 
galaxies. We obtained number--flux functions for AGNs and Galactic 
sources. The $\log N$--$\log S$ relation of AGNs (excluding blazars)
is based on 69 sources with fluxes higher than $S_{\rm
lim}=1.1\times10^{-11}$ \ergscm\ ($\sim0.8$~mCrab) in the 17--60~keV
energy band. The cumulative number--flux function of AGNs located at
Galactic latitudes $|b|>5^\circ$, where the survey is characterized by
high identification completeness, can be described by a power law with
a slope of $1.62\pm0.15$ and normalization of
$(5.7\pm0.7)\times10^{-3}$ sources per deg$^{2}$ at fluxes
$>1.43\times10^{-11}$~\ergscm\ ($>1$~mCrab). AGNs with fluxes higher
than $S_{\rm lim}$ make up $\sim1\%$ of the cosmic X-ray background at
17--60~keV. We present evidence of strong inhomogeneity in the spatial
distribution of nearby ($\lesssim 70$~Mpc) AGNs, which reflects the
large-scale structure in the local Universe.
\keywords{Surveys -- X-rays: general -- Galaxy: general -- Galaxies: Seyfert -- (Cosmology:) large-scale structure of Universe} } \maketitle

\section {Introduction}

The INTEGRAL observatory \citep{integral} has been successfully
operating in orbit since its launch in 2002. Due to the high
sensitivity and relatively good angular resolution of its instruments,
in particular the coded-mask telescope IBIS \citep{ibis}, surveying
the sky in hard X-rays is one of the primary goals of INTEGRAL. A
number of papers have reported results of deep observations of
relatively compact regions of the sky
\cite[e.g][]{revetal03a,molkov2004,krietal05,revetal06} and of 
systematic searches for sources over very large sky areas (e.g., 
\citealt{biretal06a,biretal06b}). However, until recently it was
difficult to use INTEGRAL data for source population studies, in particular
extragalactic, because the coverage of the sky remained substantially
incomplete. Therefore, in 2005--2006 we performed dedicated observations of the
previously unobserved regions of the sky and thereby completed the
most sensitive ever all-sky survey in hard X-rays. In this paper, we
present a catalog of sources detected during the all-sky survey
(Sect. \ref{sec:catalog}), discuss the number--flux relations of
Galactic and extragalactic hard X-ray sources (Sect.
\ref{subsection:lognlogs}), and investigate the spatial distribution
of local AGNs (Sect. \ref{sec:anisotropy}).

\section {Survey coverage and sensitivity}

The present survey is based on observations performed during the first
four years of the INTEGRAL mission. We used data from the
ISGRI detector of the IBIS telescope, which is well suited for
carrying out imaging surveys in hard X-rays. The coded-mask
telescope IBIS provides a wide field of view of
$28^{\circ}\times28^{\circ}$ ($9^{\circ}\times9^{\circ}$ fully coded) 
and moderate angular resolution of $12$\arcmin. The localization
accuracy of $<$2--3$\arcmin$ is sufficiently good for searches of soft
X-ray and optical counterparts and subsequent optical
identification of newly discovered hard X-ray sources.

\begin{figure}
 \includegraphics[width=0.5\textwidth]{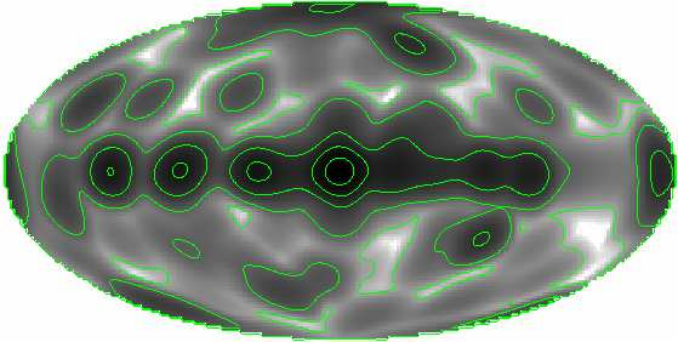}
\caption{Dead time-corrected exposure map of the survey.
The green contours represent exposure levels of 10, 150, 800 ks, $2$
and $4$~Ms.}\label{fig:expomap} 
\end{figure}

\begin{figure}
 \includegraphics[width=0.5\textwidth]{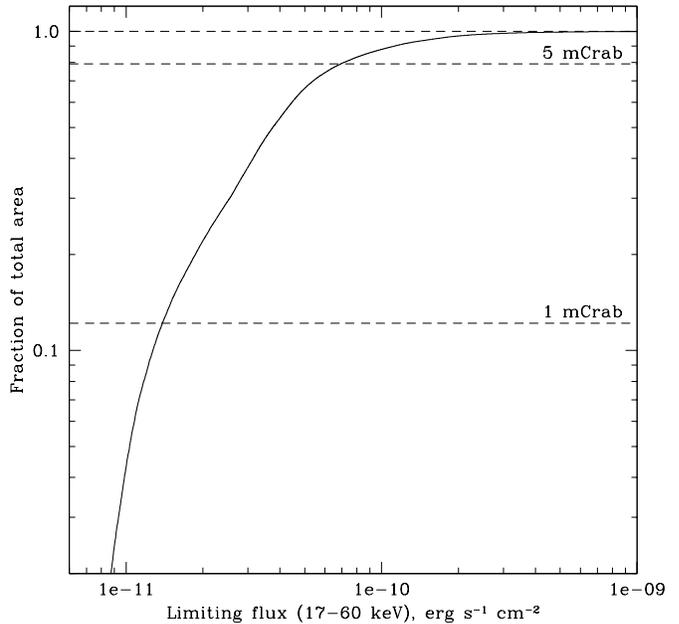}
\caption{Fraction of the sky surveyed as a function of the limiting
flux for source detection with $5\sigma$ significance.}\label{fig:area} 
\end{figure}

We used all public and proprietary data available to us. The
observations were performed during spacecraft revolutions from 25 (end
of December 2002) to 463 (June 2006). Importantly, our data set
includes the special series of thirteen 200 ks-long extragalactic
pointings (PI Churazov), which allowed us to complete the survey of the
entire sky. Figure~\ref{fig:area} shows the fraction of the
sky covered by the survey as a function of the limiting flux for
source detection with at least $5\sigma$ significance. Approximately
12\% and 80\% of the sky are covered down to 1 and 5 mCrab,
respectively. After data cleaning and dead-time correction, the total
exposure time of the survey is $\sim 33$~Ms. 

\section{Data analysis} \label{sec:analysis}

We analyzed the entire set of INTEGRAL observations at the
level of individual pointings (science windows, SCWs), which have typical
exposures of 2~ksec. For each observation, the IBIS/ISGRI raw
events list was converted to a sky image in our working energy band
(17--60 keV). The employed algorithm of image 
reconstruction was previously described by \cite{revetal04} and
\cite{krietal05}. Here we outline only those steps that are essential for the
present study.

We first accumulated raw detector images in the 17--60~keV energy
band and cleaned them from bad and noisy pixels. The reconstruction starts with
rebinning the raw detector images onto a grid with a pixel size equal to 1/3 of
the mask pixel size. This is close but not exactly 
equal to the detector pixel size. Therefore, the rebinning causes a
moderate loss in spatial resolution but enables a straightforward
application of standard coded-mask reconstruction algorithms
(e.g., \citealt{fenimor81}, \citealt{skinner87a}). Essentially, for
each sky location, the flux is calculated as the total flux in those detector
pixels that ``see'' that location through the mask minus the flux in
those detector pixels that are blocked by the mask.

The image reconstruction is based on the DLD deconvolution procedure
(see the notations in \citealt{fenimor81}), with a mask pixel
corresponding to $n\times n$ detector pixels. The original detector is
treated as $n\times n$ independent detectors, and $n\times n$
independent sky images are reconstructed and then combined into a
single image. A point source in such an image is represented by a
$n\times n$ square. In our case, this leads to the effective point
spread function (PSF) being approximately a square of $3\times3$
detector pixels, or $12'\times12'$. After summing a large number of
individual images, the 2D shape of the PSF can be well approximated
by a Gaussian with $\sigma=1.25'$.

The periodic structures in the IBIS mask cause the appearance of
parasitic peaks (''ghosts'') in the images of real sources. We used an
iterative procedure to eliminate such ghosts. For this purpose, we
used a ``current'' catalog of sources, which was renewed each time a
new bright source was detected. The image reconstruction was then
redone for each observation containing the new source using the
updated current catalog.

{\it \flushleft The case of a bright source} 

Since the pattern of the shadowgram cast by a point source 
through the mask is not ideally known, the ghost removal procedure 
is not perfect. Some photon counts can be left or oversubtracted at
certain positions on the detector. This effect is usually small but
can become significant for deep fields containing very bright sources. In
this case, characteristic ``crosses'' and ``rings'' appear around the bright
sources. The former artifact appears when the observational
program of a bright source is dominated by starring pointings
with a constant roll angle. A variable roll angle diminishes this
effect but produces concentric structures. In practice this means
that in regions with very bright sources, some source detections based
on the criterion of exceeding a reasonable threshold, say
$5\sigma$, may be false. The level at which imperfect ghost
removal starts to play a role depends on the observational pattern 
and typically corresponds to $\sim$(3--5)$\times 10^{-3}$ of the flux of
the brightest source within the IBIS field of view. Therefore, the
extracted list of excesses should always be checked ``by eye'' and cleaned
out from such characteristic series of false detections around bright
sources. It should be noted that such a cleaning introduces
``dead zones'' where a real source may be missed. We have verified that
in the worst case, the total area of such zones does not exceed
$100\textrm{ deg}^2$, which constitutes a neglible fraction of the
total area of our all-sky survey.

On applying the procedures described above, each observation
is represented by a $28^{\circ}\times28^{\circ}$ sky image with a pixel size of
4$\arcmin$. The full analyzed data set contains 23,547 such images and
comprises $\sim35$~Ms (dead time-corrected) worth 
of observations. We applied an additional filtering to the resulting
images using information about the residual (after subtraction of point
sources) $rms$ signal-to-noise variations in the
images. Those images having $rms>1.05$ were excluded from the
analysis. This resulted in an additional rejection of $\sim 7\%$ of
the pointings and finally left us with $\sim 33$~Ms worth of clean
observations. 

An analysis of mosaic images built from observations covering the whole sky
is complicated by various effects of projections onto a 2D plane. The
most significant one is distortions of the point spread function
(PSF) at positions far from the center of the projection. This leads
to uncertainties in estimating the source position and flux. To avoid
this effect, we followed an approach that had been developed for
analyzing data distributed on a sphere. Specifically, we used a number
of subroutines from the HEALPIX package (\textbf{H}ierarchical
\textbf{E}qual \textbf{A}rea 
iso\textbf{L}atitude \textbf{Pix}elization of a sphere,
\citealt{healpix}) to build a mosaiced sky. This technique provides
an equal-area pixelization of the sphere and allows us to analyze
all of the data uniformly. We produced a HEALPIX-based map of the whole
sky with 12M pixels, which corresponds to a $\sim3.4\arcmin$ size of
each sky pixel. The map was constructed by projecting individual IBIS/ISGRI
images onto the HEALPIX all-sky frame. 

\subsection{Detection of sources}

We searched for sources on three time scales -- in 
individual SCWs ($\sim$2~ks exposure time, typical sensitivity
20--30~mCrab), on images integrated over individual satellite orbits 
($\sim$200~ks, $\sim4$ mCrab), and on the time-averaged all-sky map
($\sim$33~Ms).

\begin{figure}
 \includegraphics[width=0.5\textwidth]{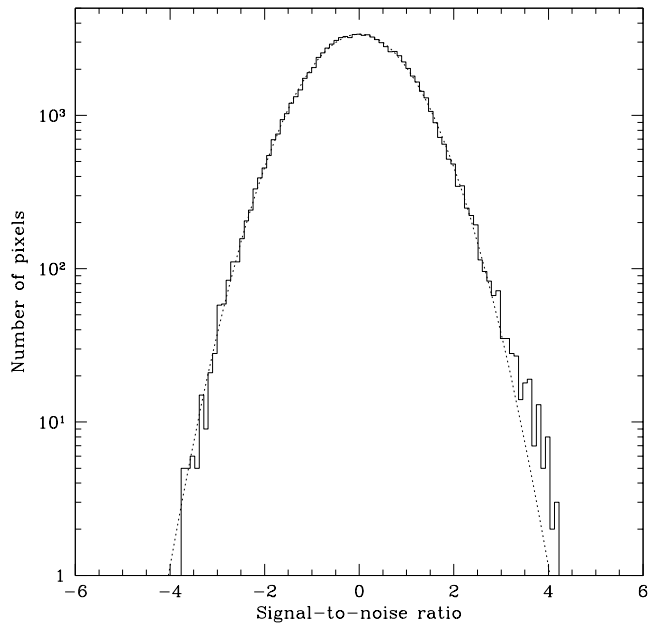}
\caption{Signal-to-noise ratio distribution (solid histogram) of the number of
pixels in a $20^{\circ}\times20^{\circ}$ projection image of the sky
around the position $15h$~$28m$~$00.0s$,~$-28^{\circ}$~$00m$~$00s$
(J2000). The dotted line represents the normal distribution with unit
variance and zero mean.}
\label{fig:rms} 
\end{figure}

The detection of sources was performed using $20^\circ \times 20^\circ$
projections of the HEALPIX all-sky image. Since adjacent pixels in
these images (pixel size $\sim 4\arcmin$) are smaller
that the instrumental PSF, we convolved the images 
with a Gaussian filter that mimics the effective instrumental PSF
($\sigma=1.25'$).

The signal-to-noise ratio distribution of pixels is dominated by the
statistical noise and can be described by a Gaussian. In those sky
regions that contain very bright point sources, the $rms$ scatter of
signal-to-noise ratios increases. However, this mainly occurs due to
the presence of unsubtracted ghosts, which can be found by eye and removed.

There are small misalignments between the mask grid and the detector
pixels, which cause imperfect redistribution of counts over the
mask-aligned detector grid. However, this effect is minor, because the
correlation length of the noise in the projected images is smaller
than the size of the applied Gaussian filter. As a result, the
originally measured variance of the distribution of pixel
significances is slightly less than unity. To correct for this, we
implemented a uniform rescaling of the error map for each sky
projection. The correction factor was determined from each HEALPIX
projection image and found to vary between 0.9 and 1.0. For sky
fields containing a large number of point sources or for images of
poor quality, the mentioned effect becomes less important than
systematic biases introduced by the image 
reconstruction algorithm (Sect. \ref{sec:analysis}). In such cases,
the correction factor cannot be properly estimated, and so we adopted
it to be not greater than unity. In Fig. \ref{fig:rms} we show the
signal-to-noise ratio distribution of the pixels of a
$20^{\circ}\times20^{\circ}$ corrected sky map centered on some
arbitrary position. This distribution (except for the positive bright
tail produced by real sources) can be described by the normal
distribution with unit variance and zero mean. 

We next specified a detection threshold in units of this clean
``sigma'' to search for sources in images. The lists of source
candidates were then cleaned by eye from excesses forming
characteristic patterns around bright sources.

Taking into account the IBIS/ISGRI angular resolution, the
all-sky map and maps accumulated during single revolutions contain
$\sim 10^6$ and $3\times10^7$ statistically independent pixels,
respectively. We adopted the corresponding detection thresholds of
$(S/N)_{\rm lim}>5\sigma$ and $(S/N)_{\rm lim}>5.5\sigma$ to ensure
that the final catalog contains less than 1--2 spurious sources.

Apart from the main search described above, we detected several
sources in special extended series of observations such as the deep 
surveys of the Sagittarius and Crux spiral arm tangent regions
\citep{molkov2004, revetal06}. Some of these sources fall below our
detection threshold ($5\sigma$) on the all time-averaged map, which
probably indicates their strong variability or transient
nature. 

We emphasize that for statistical studies only those sources from the
catalog that have time-averaged statistical significance higher than
$5\sigma$ (see the flux column in Table~\ref{tab:catalog} below)
should be used.

\subsection{Localization accuracy}

We determined the positions of sources by fitting the centroid of a 2D gaussian
($\sigma=1.25'$) to the peak of the PSF-convolved source image. To
estimate the accuracy of this method, we   
built the distribution of deviations of the measured positions of
sources with known cataloged locations for a large number of INTEGRAL
observations. The positional accuracy of sources detected by
IBIS/ISGRI depends on the source significance \citep{gros2003,biretal06a}. The 
estimated $68\%$ confidence intervals for sources detected at 5--6,
10, and $>20\sigma$ are $2.1$\arcmin, $1.5$\arcmin, and $<0.8$\arcmin,
respectively. 

\section{Catalog}
\label{sec:catalog}

We detected a total of 400 sources in the 17--60 keV energy
band over the whole sky. The full list of sources is presented in
Table \ref{tab:catalog}, and its content is described below.

{\it Column (1)} -- source number in the catalog.

{\it Column (2)} -- source name. For sources whose nature was known 
before their detection by INTEGRAL, their common names are given. Sources
discovered by INTEGRAL or those whose nature was established thanks to
INTEGRAL are named ``IGR''

{\it Columns (3,4)} -- source Equatorial (J2000) coordinates.

{\it Column (5)} -- time-averaged source flux in mCrab units. A flux
of $1$~mCrab corresponds to $1.43\times10^{-11}$ \flux ~for a source
with a Crab-like spectrum.

{\it Column (6)} -- general astrophysical type of the object: LMXB
(HMXB) -- low- (high-) mass X-ray binary, AGN -- active galactic
nucleus, SNR/PWN -- supernova remnant, CV -- cataclysmic variable, PSR
-- isolated pulsar or pulsar wind nebula, SGR -- soft gamma repeater,
RS CVn -- coronally active binary star, SymbStar -- symbiotic star, Cluster --
cluster of galaxies.

{\it Column (7)} -- additional notes and/or alternative source names.  

{\it Column (8)} -- references. These are mainly provided for new
sources and are related to their discovery and/or nature.

We note that \cite{biretal06b} have recently performed a similar hard X-ray
survey using INTEGRAL/IBIS/ISGRI data. The catalog of these authors
contains 421 sources detected in five energy bands spanning
18--100~keV. Although a detailed comparison of this catalog with ours
goes beyond the scope of the present paper, we may mention some important 
differences: (i) our dataset covers a number of extragalactic regions
not covered by the survey of \cite{biretal06b}, on the other hand
their dataset contains a considerable amount of data not available to us; (ii) 
\cite{biretal06b} used the standard INTEGRAL OSA software, whereas  
we used a software developed at the  Space Research Institute (Moscow,
Russia); (iii) the detection criterion adopted in our work allows not
more than 1--2 spurious sources to be present in the whole catalog, while
the catalog of \cite{biretal06b} may, by construction, contain
considerably more spurious sources. 

\subsection{Some peculiar sources}

\emph{Galactic Center source IGR J17456$-$2901}

The sky density of hard X-ray sources is not very
high -- in general $N(>1 \textrm{mCrab})<0.1$ deg$^{-2}$, therefore
the angular resolution of the IBIS telescope
($\sim12\arcmin$) is usually sufficient to prevent source confusion. The
only exceptional region is the Galactic center: in the close
vicinity (within $1^\circ$) of Sgr A$^*$, IBIS sees 10
sources.

At the position of Sgr A$^*$ there is an additional hard X-ray excess, IGR
J17456$-$2901. This source was originally reported by 
\cite{revetal04c} and erroneously associated with the X-ray
burster AX J17456$-$2901. Subsequent studies of this source
demonstrated that it is extended \citep{neronov05,belanger05} and is probably
the superposition of a large number of faint point sources located in 
the Galactic nuclear stellar cluster \citep{krivetal05b}. Note that,
as is shown by \cite{belanger05}, at higher energies ($\sim$ 70--100~keV),
the position of the centroid of the excess is displaced with respect to
Sgr A$^*$, indicating that the nature of the high-energy source may be
different from that of the 17--60 keV emission. 

\emph{\flushleft RX J1713.7$-$3946}

Since IBIS is a coded-mask telescope, it is not well suited for
studying sources more extended than its angular resolution. However,
if a source is only somewhat larger than the instrumental PSF, it is 
possible to obtain some limited information about the spatial
structure of the source \citep[see e.g.][]{renaud2006a,renaud2006b}.

In particular, in our catalog there are four extended sources that are not much
larger than the IBIS PSF ($\sim12\arcmin$): three
clusters of galaxies (Oph, Perseus, and Coma) and the supernova 
remnant RX J1713.7$-$3946. 

\begin{figure}
 \includegraphics[width=0.5\textwidth]{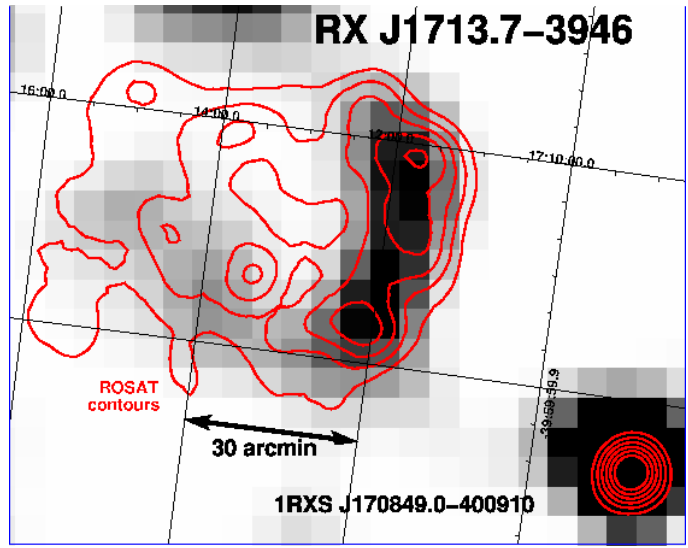}
\caption{INTEGRAL/IBIS hard X-ray (17--60 keV) image of the
supernova remnant RX J1713.7$-$3946. The gray scale on the map is
proportional to hard X-ray flux. The map obtained by ROSAT in the soft
X-ray (0.5--2.5 keV) band is shown by contours.}
\label{fig:rxj1713}
\end{figure}

It is clearly seen (Fig.~\ref{fig:rxj1713}) that the supernova remnant 
RX~J1713.7$-$3946 exhibits clear extended structure
(visible size in hard X-rays $\sim24$\arcmin). The
significance of the hard X-ray detection varies along the extended structure
between 4$\sigma$ and 5$\sigma$ (the statistical significance of the total
extended emission is $>10\sigma$). The total
exposure for this region is 5.3~Ms (dead time-corrected). To
test the stability of the apparent spatial feature, we split the
entire period of observations into four intervals and examined them
individually. The extended structure is clearly present in each image and
looks stable against the background of variable noise.

The supernova remnant RX~J1713.7$-$3946 was discovered in soft X-rays
during the ROSAT all-sky survey \citep{pfeffermann1996}. An extended elliptical structure was
found with a maximum extent of 70\arcmin ~(see the red contours in 
Fig. \ref{fig:rxj1713}). The non-detection of emission lines in the
X-ray spectrum
of RX~J1713.7$-$3946 (ASCA, \citealt{koyama1997,slane1999}) was regarded
as an indication that the observed X-rays is non-thermal emission
from an expanding shell. 

Recently, very high-energy (VHE) gamma-ray emission was discovered from the
remnant by the H.E.S.S. experiment \citep{aharonian2006}. The spatial
correlation of the VHE emission intensity with the X-ray morphology
confirms that cosmic-ray particles are being accelerated in the shell.

Here we report a detection of RX~J1713.7$-$3946 in hard X-rays. The
hard X-ray emission is probably synchrotron emission of 100-TeV
electrons accelerated in the shell \citep{koyama1995}. 

\section{Extragalactic sources -- AGNs}
\label{subsection:lognlogs}

All sources in our catalog can be separated into two main classes
-- galactic and extragalactic (mainly AGNs). Under the assumption that
AGNs are uniformly distributed over the sky (which is only a crude 
approximation of the real situation, see 
Sect.~\ref{sec:anisotropy}), we can construct the deepest ever
number-flux function of hard X-ray emitting AGNs.

The catalog contains 131 objects\footnote{Three additional sources:
IGR~J02466$-$4222, IGR J02524$-$0829, and IGR J18578$-$3405 have a suspected
AGN origin.}  identified as AGNs (see also
\citealt{sazetal07} \footnote{After publication of the INTEGRAL AGN catalog
by \citet{sazetal07}, four sources have been added 
to the AGN list: IGR~J18249$-$3243 \citep{basetal06}, ESO~005-G004,
IGR J14561$-$3738, and SWIFT 0920.8$-$0805}). Of these, 94 have
statistical significance higher than 5$\sigma$ on the time-averaged
map, including 86 emission-line AGNs (non-blazars) and 8
blazars. There are also 40 unidentified sources detected on the
average map. The relative fraction of unidentified sources is much
smaller for the extragalactic sky ($|b|>5^{\circ}$) than for the whole
sky: the corresponding numbers of non-blazar AGNs and unidentified
sources are 69 and 6.

\begin{figure}
\includegraphics[width=0.5\textwidth]{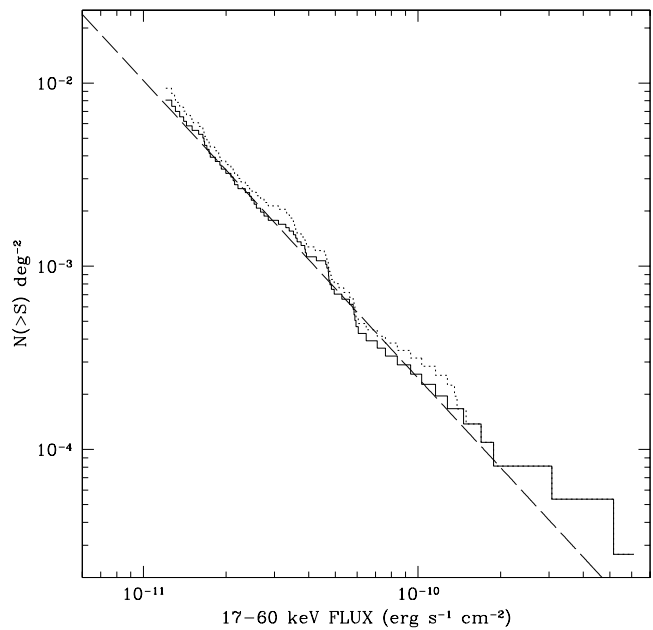}
\caption{Cumulative $\log N$--$\log S$ distribution of non-blazar AGNs
in the energy band 17--60~keV obtained in the extragalactic sky
($|b|>5^\circ$, solid histogram). The best-fitting power law with a
slope of $1.62\pm0.15$ and normalization of
$(5.7\pm0.7)\times10^{-3}$~deg$^{-2}$ at a flux of 1~mCrab is shown by
the dashed line. The dotted curve represents the $\log N$--$\log S$
distribution of all extragalactic sources including blazars and
clusters of galaxies (except for unidentified sources).}
\label{fig:lognlogs}
\end{figure}

Since INTEGRAL observations cover the sky inhomogeneously, in
constructing number-flux functions we should take the sensitivity map
into account. In Fig. \ref{fig:lognlogs} we show the 
cumulative $\log N$--$\log S$ distribution of non-blazar AGNs derived 
at $|b|>5^\circ$ (excluding the 7 unidentified sources). It can
be well fit by a power law: $N(>S)=A S^{-\alpha}$. Using a
maximum-likelihood estimator \citep[see e.g.][]{crawford1970}, we
determined the best-fit values of the slope and normalization:
$\alpha=1.62\pm0.15$ and $A=(5.7\pm0.7)\times10^{-3}$ deg$^{-2}$ at
$S=1$~mCrab. This implies that AGNs with fluxes exceeding  
our effective threshold $S_{\rm lim}=0.8$~mCrab account for $\sim1\%$ of
the intensity of the cosmic X-ray background in the 17--60~keV band, 
which was recently re-measured by INTEGRAL (Churazov et al. 2007).


We previously \citep{krietal05} constructed a number--flux relation of
extragalactic sources in a relatively small region of the sky
($45^{\circ}\times45^{\circ}$) around the Coma cluster of
galaxies. The deep ($\sim500$~ks) observations of the Coma were used
to study a sample of 12 serendepitously detected sources in that
field. After correcting for the expected number of false detections
and fitting the resulting $\log N$--$\log S$ relation by a power law
with the Euclidean-geometry slope of $3/2$, the surface density of hard X-ray
sources above a 20--50 keV flux threshold of $10^{-11}$~\ergscm\
($\sim1$~mCrab) was found to be
$(1.4\pm0.5)\times10^{-2}$~deg$^{-2}$. This value is significantly 
higher than the average surface number density of AGNs in the $|b|>5^\circ$
sky, determined above, which probably reflects the large-scale
overdensity of galaxies in the general direction of the Coma cluster (see
Sect. \ref{sec:anisotropy}).

\begin{figure}
\includegraphics[width=0.5\textwidth]{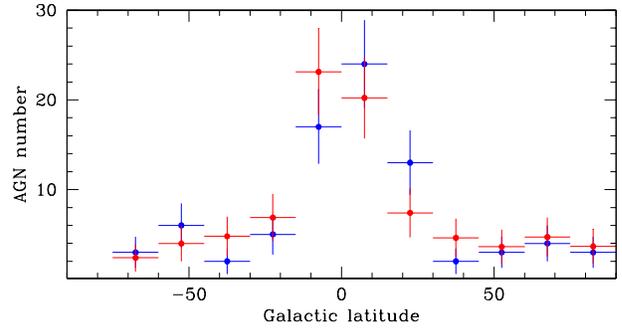}
\caption{Expected numbers of AGNs (red) in $15^{\circ}$-wide Galactic-latitude 
strips (integrated over all Galactic longitudes) and the
corresponding numbers of actually detected and identified AGNs
(blue). The involved statistical uncertainties are shown by error bars.
}\label{fig:latnumber} 
\end{figure}

Using the derived $\log N$--$\log S$ distribution, we can compare the
numbers of AGNs detected during the survey in different parts of the
sky with the numbers expected under the assumption of uniform spatial
distribution of sources. We find good agreement between these
numbers within the statistical errors for $15^\circ\times 360^\circ$
strips cut parallel to the Galactic plane
(Fig. \ref{fig:latnumber}). Even in the Galactic plane region
($|b|<5^{\circ}$), which was excluded from our calculation of the
number-flux function, the expected number of AGNs exceeding the
detection threshold (18.4) is compatible with the number of detected
and identified AGNs (16). This suggests that most of the unidentified 
sources in the Galactic plane region are of Galactic, rather than
extragalactic origin. This tentative conclusion of course rests on our
assumption that AGNs are distributed uniformly on very large
scales over the sky, which is in fact approximately true only for
relatively distant objects ($D\gtrsim 70$~Mpc, see the next Section)
and much less so for more nearby AGNs, which constitute approximately
half of our sample.

If we now include all the identified AGNs detected in the Galactic
plane region into the calculation of the AGN number--flux function
(thus increasing the total number of non-blazar AGNs to 86 and
extending the calculation to the whole sky), we find
$\alpha=1.50\pm0.13$ and $A=(5.4\pm0.6)\times10^{-3}$ deg$^{-2}$ at 1
mCrab, i.e. virtually the same values as for the
$|b|>5^\circ$ sky.

\section {Signatures of large-scale structure in the population of
hard X-ray AGNs}
\label{sec:anisotropy}

It is now widely accepted that practically every galaxy in the local 
Universe has a supermassive black hole and some of these black holes
are AGNs with widely ranging luminosities (see e.g.,
\citealt{richstone1998,kormendy2001}, for a review). Therefore, it is reasonable to assume that
the space density of X-ray emitting AGNs is proportional to that of
normal galaxies.

The spatial distribution of galaxies in the local Universe is
inhomogeneous. The gravitational attraction of matter in the Universe
has formed different structures with sizes\footnote{Hereafter we adopt
$H_0=73$~km~s$^{-1}$~Mpc$^{-1}$.} up to $\sim 100$~Mpc. On larger
scales, matter is distributed more or less uniformly, whereas on smaller scales
there is strong inhomogeneity. The contrast in matter density
between galaxy concentrations and voids can reach an order of magnitude
and more \citep[see e.g.][]{rees80,davis83,bahcall86}. As our sample
of hard X-ray emitting AGNs mostly probes the nearby Universe out to
distances $\sim 200$~Mpc, we have the possibility to see
similarly strong inhomogeneities in the distribution of nearby
AGNs.

To this end, we estimated the space densities of AGNs in different
directions of the sky. Due to the relatively small size of our sample, we
assumed that the AGN number density is constant along a given line of
sight while the shape of the AGN luminosity function is invariant in
the local Universe. We adopted this shape from \citet{sazetal07}, who
calculated the all-sky average hard X-ray luminosity function using the same
sample of AGNs as in the present study.

Under these assumptions and using the sensitivity map of the survey, we 
determined the normalization of the luminosity function within spherical
cones drawn around multiple directions in the sky through comparison of the
expected and measured numbers of AGNs in these cones. We adopted
the half-opening angle of the cones to be $\theta=45^\circ$ in order
to achieve reasonably good angular resolution of the resulting map and
still have a significant number of AGNs in each cone. To optimize our
sensitivity to anisotropies in the spatial distribution of AGNs, we
restricted ourselves to distances $< 70$~Mpc, at which maximal
contrasts in galaxy numbers are expected \cite[see e.g.][]{rowanetal00}.

\begin{figure}[h]
 \includegraphics[width=\columnwidth]{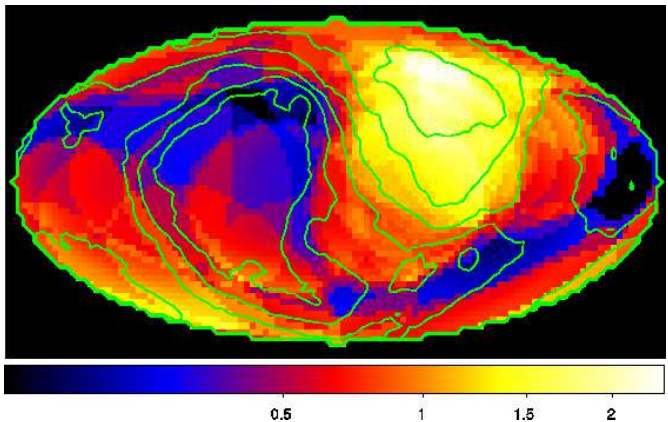}
\caption{2D-map of the AGN number density in the local Universe. This
map was constructed in Galactic coordinates using the sample of
39 AGNs located at distances $D<70$~Mpc. Shown for each pixel is the
estimated normalization of the AGN luminosity function within a
spherical cone with a half-opening angle of $45^{\circ}$ around that
direction. The density is given in units of
$2\times10^{-4}$~Mpc$^{-3}$ at luminosities higher than
$10^{42}$~erg~s$^{-1}$ (17--60~keV), which is the average local
density of AGNs
\citep{sazetal07}. Green contours show the surface number density of galaxies 
detected during the IRAS PSCz survey at distances $D<70$~Mpc.} 
\label{fig:map1}
\end{figure}

\begin{figure}[h]
 \includegraphics[width=\columnwidth]{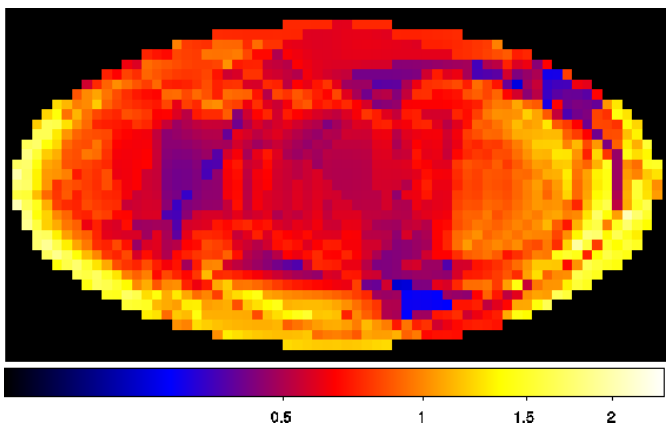}
\caption{2D-map of the number density of AGNs at distances $D>70$~Mpc.
See Fig. \ref{fig:map1} for the description. This map is based on 40 obejcts.}
\label{fig:map2}
\end{figure}

In Sect. \ref{sec:analysis} we demonstrated that most of the extragalactic
objects detectable by INTEGRAL in the Galactic plane region are
probably already identified, hence we can use for our analysis the
all-sky sample of AGNs.

The resulting map of the number density of nearby AGNs over the sky is
shown in Fig. \ref{fig:map1}. The density is given in units of
$2\times10^{-4}$~Mpc$^{-3}$ at luminosities higher than
$10^{42}$~erg~s$^{-1}$ (17--60~keV), which is approximately the
average local density of AGNs \citep{sazetal07}. Prominent large-scale
concentrations of AGNs can be seen in the  northern (to the right)
and southern (to the left) Galactic hemispheres. We can assess the
statistical significance of these anisotropies by considering the
density measurements in statistically independent regions of the
sky. Given the relatively large opening angle of our sampling cone,
there are only 12 such independent directions on the whole sky. Since
in our case the estimated density values are strongly affected by
Poisson statistics, we constructed a maximum likelihood estimator in
the form $L=\sum{\log P_x(n)}$, where $P_x(n)$ is the probability to
detect $n$ sources in our detection cone if the expected number of
sources is $x$. The probability that the measured AGN number density
distribution is a statistical realization of a homogeneous
distribution is $8.7\times10^{-4}$.

If we take into account that the high and low AGN number density
regions well correlate with known (from surveys of infrared 
galaxies, e.g. \citealt{rowanetal00}, and X-ray clusters of galaxies, e.g.
\citealt{kocetal06}) over- and under-dense regions in the local
Universe, respectively, the statistical significance of the found
spatial inhomogeneity will be much higher.

The discovered anisotropy of AGNs agrees well with the known
distribution of matter in the local Universe. The large-scale feature
in the northern Galactic hemisphere is consistent with the position of the
highest mass concentrations in the local Universe: the nearby Virgo
cluster ($\sim18$~Mpc, $\sim1.2\times10^{15}\textrm{M}_{\sun}$
\citealt{fouque01,tonry00}) and the more distant and massive Great Attractor
($\sim65$~Mpc, (1--5)$\times10^{16}\textrm{M}_{\sun}$,
\citealt{lynden1988,tonry00}). The southern structure is consistent
with the Perseus-Pisces supercluster ($\sim50$~Mpc,
(7--9)$\times10^{15}\textrm{M}_{\sun}$,
\citealt{hanski01}).

To better demonstrate the similarity between the distributions of hard
X-ray emitting AGNs and matter over the sky, we used the IRAS PSC
redshift survey \citep{saunders00}. We selected galaxies located at
distances $<70$~Mpc and have far-infrared flux $S_{\rm 60\mu
m}>1$~Jy. The IRAS PSCz survey covers approximately 83\% of the sky
due to presence of so-called Zone of avoidance, sky region to the
north and to the south from the Galactic equator where Galaxy obcures
the IR emission. In order to fill this gap during the construction of
the map of densities of IR galaxies we assumed that the number density
of galaxies hidden behind the Galactic plane (10 degrees to the north
and to the south of the equator) is constant and equals to the all-sky
average value. Contours of the number density map of IRAS galaxies are
shown in Fig.~\ref{fig:map1}. We emphasize that this comparison is
rather approximate and only reflects the distribution of the projected
mass. A detailed study of the correlation between the spatial
distributions of hard X-ray emitting AGNs and matter in the local
Universe will be presented elsewhere.

\begin{figure}
 \includegraphics[width=0.5\textwidth]{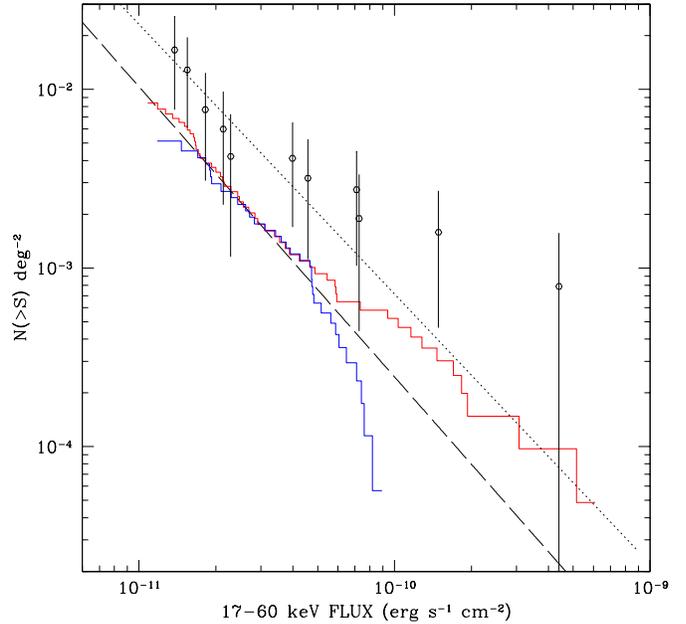}
\caption{
Number-flux functions of extragalactic sources (excluding blasars
and clusters of galaxies) measured in different sky regions. The
$\log N$--$\log S$ relations for sources in two hemispheres, in the
direction of the IRAS dipole $l=268^\circ$, $b=27^\circ$
\citep{lahav1988} and in the opposite direction, are shown by the red
and blue histogram, respectively. The dashed line represents the
best-fitting $\log N$--$\log S$ distribution for the entire sky
excluding the Galactic plane ($|b|>5^{\circ}$, see
Fig. \ref{fig:lognlogs}). The open circles with error bars represent
the number-flux relation of extragalactic sources obtained in the
$45^{\circ}\times45^{\circ}$ region around the Coma cluster
\citep{krietal05} in the 20--50~$\textrm{ keV}$ energy band. The flux
was converted to the 17--60~$\textrm{ keV}$ energy band assuming Crab
spectrum.  The dotted line shows the corresponding $N\propto S^{-3/2}$
fit.}
\label{fig:lognlogsdipole} 
\end{figure}

It is obvious from the above discussion that any estimate of the AGN
surface number density based on a small area of the sky may be
significantly biased. This was apparently the case for our survey in
the $\sim45^{\circ}\times45^{\circ}$ region around the Coma cluster
\citep{krietal05}, where a high surface number density of relatively
nearby ($D\lesssim 70$~Mpc) AGNs was found (see
Sect.~\ref{subsection:lognlogs}). Indeed, the observed region is
located approximately in the direction of the prominent large-scale
structure in the northern Galactic hemisphere (see
Fig.~\ref{fig:map1}). To better demonstrate the strong contrast in the
distribution of AGNs over the sky, we calculated the
sensitivity-corrected $\log N$--$\log S$ distributions of AGNs in the
two hemispheres defined by the direction of motion of the Local Group:
$l=268^\circ,b=27^\circ$, as measured by IRAS \citep{lahav1988}. These
distributions are shown in Fig.~\ref{fig:lognlogsdipole}. The counts
of bright sources ($S\ga10^{-10}$~\ergscm), for which our survey is
almost 100\% complete in both hemispheres, exhibit a contrast as high
as $11:1$.

In addition, we explicitly calculated the AGN luminosity functions in
both hemispheres defined above and found their shapes to be consistent
with that of the all-sky average luminosity function determined by
\cite{sazetal07}. This confirms that the found anisotropy mostly
reflects the inhomogeneous distribution of matter in the local Universe
rather than generic variations of the AGN luminosity function.

Figure~\ref{fig:lognlogsdipole} demonstrates that most of the observed
anisotropy in the distribution of AGNs over the sky is due to closest
(brightest) AGNs. Inclusion of sources located at progressively larger
distances is expected to decrease the surface number density
variations over the sky \citep{rowanetal00}. To demonstrate
this, we built a map of the number density of AGNs located 
at $D>70$~Mpc (Fig.~\ref{fig:map2}), which should be compared with
that for nearby AGNs ($D<70$~Mpc, Fig. \ref{fig:map1}). It
can be seen that the more distant AGNs are distributed more uniformly
across the sky, although their distribution is still only marginally
consistent with a homogeneous one: the corresponding
probablity is $10^{-2}$. 

The open circles in Fig. \ref{fig:lognlogsdipole} show the 
$\log N$--$\log S$ distribution previously obtained in the Coma
region. This number-flux relation lies higher than the all-sky average
due to the overdensity of galaxies in this relatively small region of
the sky (1,243~sq.~deg), including two bright AGNs, NGC 4151 and NGC
4388. As a result, the resolved fraction of the CXB in this region is
also high, $\sim 3$\%. As was already mentioned above, the resolved
fraction of the CXB for the whole sky is only $\sim 1$\%. This last value
is consistent with that reported by \cite{beckmann06} based on their
analysis of INTEGRAL observations covering 25,000~sq.~deg of the sky. 

\section{Galactic sources}
\begin{figure}
\includegraphics[width=0.5\textwidth]{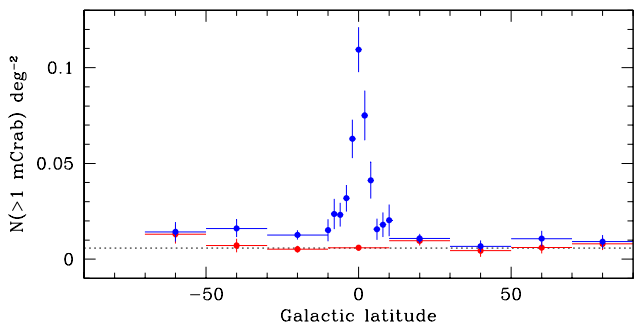}
\caption{Surface number density of ($>5\sigma$) sources with flux
$>1$~mCrab as a function of Galactic latitude. Blue and red points
represent all and (identified) extragalactic sources,
respectively. The dashed line represents the normalization of the
all-sky extragalactic $\log N$--$\log S$ function.}
\label{fig:latnorma2} 
\end{figure}

The presence of a large number of Galactic sources in the INTEGRAL
all-sky catalog is obvious from the large overdensity of sources near
the Galactic plane. For demonstration we built sensitivity-corrected
cumulative number-flux functions for all 
($>5\sigma$) sources in $15^{\circ}$ Galactic-latitude bins. The
derived surface number density of sources as a function of Galactic
latitude is shown in Fig. \ref{fig:latnorma2}.

\begin{figure}
 \includegraphics[width=0.5\textwidth]{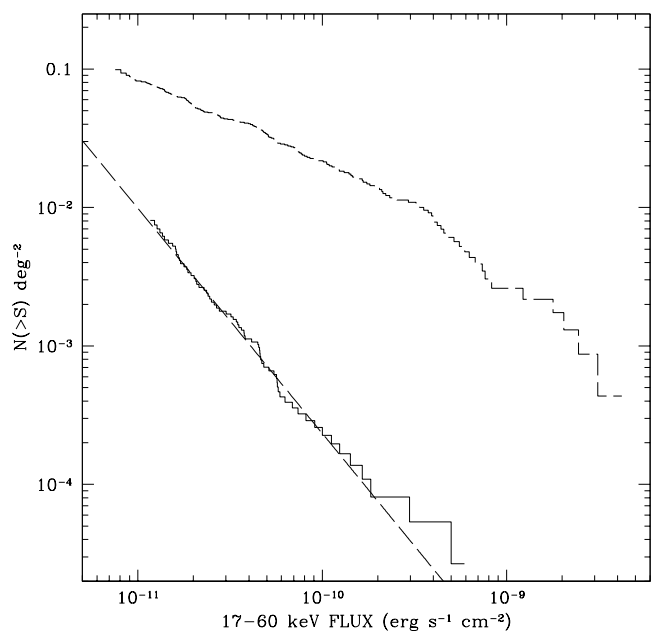}
\caption{Cumulative $\log N$--$\log S$ distribution in the energy band
17--60~keV of all sources in the Galactic plane region
($|b|<5^{\circ}$, dashed histogram) in comparison with that of AGNs at
high latitudes ($|b|>5^{\circ}$, solid histogram). The dashed line
represents the best-fitting power law to the number-flux relation of AGNs.}
\label{fig:lognlogsgplane}
\end{figure}

The vast majority of sources in the Galactic plane are low- and
high-mass X-ray binaries ($>70\%$ in total, excluding unidentified
sources). The number-flux function of all sources at 
$|b|<5^\circ$ (Fig. \ref{fig:lognlogsgplane}), is much flatter
than that of extragalactic sources at $|b|>5^\circ$ \citep{uhuru} and
reflects the luminosity functions of the dominant Galactic source
populations \citep[see e.g.][]{grimm02}. A detailed study of Galactic sources
based on the INTEGRAL all-sky survey will be presented elsewhere.

\section{Conclusion}
We have presented the all-sky hard X-ray (17--60 keV) survey performed by the
IBIS coded-mask telescope of the INTEGRAL observatory. The total
dead-time corrected exposure of the survey is $33$~Ms. $12\%$ and
$80\%$ of the whole sky have been covered down to limiting fluxes of $1$
and $5$~mCrab, respectively. The survey allowed us for the first time
to study the population of hard X-ray sources in an unbiased manner,
without strong influence of limited sky coverage.

Our catalog contains 400 detected sources. Of these, 213 are of Galactic
origin (87 LMXBs, 74 HMXBs, 21 CVs, 6 active stars, and other types)
and 136 are extragalactic, including 131 AGNs, 3 AGN candidates, and 3
clusters of galaxies. There remain 48 (including 40 detected with more
than 5$\sigma$ significance on the time-averaged 
map) unidentified sources, only 6 of which are located in the
extragalactic sky ($|b|>5^\circ$).  The catalog includes 137 sources
discovered by INTEGRAL, 15 of which are reported here for the first
time.

A number of sources detected by INTEGRAL/IBIS have counterparts 
in the TeV energy band. In this paper we for the first time presented the map
of extended hard X-ray (17--60~keV) emission of the supernova remnant
RX J1713.7$-$3946. 

We explored the spatial distribution of AGNs in the local
Universe. The cumulative $\log N$--$\log S$ function of AGNs derived
away from the Galactic plane (at $|b|>5^\circ$) can 
be well fit by a power law $N(>S)=(5.7\pm0.7)\times10^{-3}
S^{-1.62\pm0.15}$~deg$^{-2}$ (fluxes in mCrab units). This implies
that $\sim1$\% of the CXB at 17--60~keV is directly resolved by INTEGRAL.

We demonstrated that local ($\lesssim 70$ Mpc) AGNs are inhomogeneously
distributed in space, largely following the large-scale
structure. In particular, significant concentrations of AGNs were found
in the regions of the sky around the Virgo cluster, the Great
Attractor, and the Perseus-Picses supercluster.

\begin{acknowledgements}
The INTEGRAL data used here were obtained from the European and
Russian INTEGRAL Science Data Centers. The work was supported by the
President of the Russian Federation (through the program of support
of leading scientific schools, project NSH-1100.2006.2), by the
Presidium of the Russian Academy of Sciences/RAS (the program
``Origin and evolution of stars and galaxies''), by the Division of
Physical Sciences of the RAS (the program ``Extended objects in the
Universe''), and by the Russian Basic Research Foundation (projects
05-02-16540 and 04-02-17276). AL acknowledges the support from the
Russian Science Support Foundation.
\end{acknowledgements}

\footnotesize
\section{References for the catalog}
(1) \cite{donetal05},
(2) \cite{sazetal05},
(3) \cite{masetal04},
(4) \cite{masetal06c},
(5) \cite{brandt05},
(6) \cite{sunyaev03a},
(7) \cite{masetal06e},
(8) \cite{masetal06b},
(9) \cite{basetal06},
(10) \cite{sguetal06},
(11) \cite{lutetal04a},
(12) \cite{lubietal05},
(13) \cite{denhar2004a},
(14) \cite{grebetal04b},
(15) \cite{smietal06},
(16) \cite{kretetal04},
(17) \cite{cheetal05a},
(18) \cite{intzand04},
(19) \cite{rodetal04},
(20) \cite{moretal06},
(21) \cite{revetal04B2},
(22) \cite{kuulk2006},
(23) \cite{revetal06},
(24) \cite{lutrev03},
(25) \cite{lutetal05a},
(26) \cite{bykov2004},
(27) \cite{basetal05},
(29) \cite{pavetal92},
(30) \cite{lutetal03c},
(31) \cite{intzand05},
(32) \cite{biketal06a},
(33) \cite{negsmi06},
(34) \cite{prodetal03},
(35) \cite{barlow2005},
(36) \cite{barlow2006},
(37) \cite{waletal06},
(38) \cite{molkov2004},
(39) \cite{bykov2006},
(40) \cite{liuetal00},
(41) \cite{liuetal01},
(42) \cite{rodetal05},
(43) \cite{denhar2005},
(44) \cite{biretal06a},
(45) \cite{denhar2004b},
(46) \cite{ubeetal05},
(47) \cite{maletal05},
(48) \cite{aleetal95},
(49) \cite{cornalisse06},
(50) \cite{biketal06b},
(51) \cite{cheetal05b},
(52) \cite{reig2005},
(53) \cite{hannetal03},
(54) \cite{sunyaev03b},
(55) \cite{bureetal06b},
(56) \cite{neguetal05},
(57) \cite{keeketal06},
(58) \cite{krivetal05b},
(59) \cite{grebetal04a},
(60) \cite{bureetal06a},
(61) \cite{grebetal05b},
(62) \cite{kuietal06},
(63) \cite{masetal06a},
(64) \cite{lutetal04b},
(65) \cite{tueletal05},
(66) \cite{halp2006},
(67) \cite{masetal06d},
(68) \cite{masetal05},
(69) \cite{karasev07},
(70) \cite{brandt06},
(71) \cite{cheretal03},
(72) \cite{gaenetal05},
(73) \cite{turletal05},
(74) \cite{grebetal05a},
(75) \cite{gotz2006},
(76) \cite{krivetal06},
(77) \cite{sazetal07},
(78) \cite{reaetal06},
(79) \cite{belanger05},
(80) \cite{palmetal05},
(81) \cite{torretal05},
(82) \cite{cheletal06},
(83) \cite{tomsick2006},
(84) \cite{revetal03a},
(85) \cite{courv2003},
(86) \cite{walter2004},
(87) \cite{tomsick2003},
(88) \cite{revetal03b},
(89) \cite{patel2004},
(90) \cite{boda2006},
(91) \cite{tomsick2004},
(92) \cite{molkov2003},
(93) \cite{lutetal05b},
(94) \cite{kuulk2003},
(95) \cite{revetal03c},
(96) \cite{lutetal03a},
(97) \cite{lutetal05c},
(98) \cite{revetal04c},
(99) \cite{augetal03},
(100) \cite{lutetal03b}
(101) Sazonov et al., in preparation.
\normalsize

\bibliography{ridge,integral}

\begin{thebibliography}{}

\bibitem[Aharonian et al.(2006)]{aharonian2006} Aharonian, F., et al.\ 2006,
\aap, 449, 223

\bibitem[Aleksandrovich et al.(1995)]{aleetal95} Aleksandrovich,
N.~L., Aref'ev, V.~A., Borozdin, K.~N., Syunyaev, R.~A., \& Skinner, G.~K.\
1995, Astronomy Letters, 21, 431

\bibitem[Augello et al.(2003)]{augetal03} Augello, G., Iaria, R., Robba, N.,
et al.\ 2003, \apj, 596, L63

\bibitem[\protect\citeauthoryear{Bahcall \&
Burgett}{1986}]{bahcall86} Bahcall N.~A., Burgett W.~S., 1986, ApJ,
300, L35

\bibitem[Barlow et al.(2005)]{barlow2005} Barlow, E.~J., Bird, A., Clark, D., et al.\
2005, \aap, 437, L27

\bibitem[Barlow et al.(2006)]{barlow2006} Barlow, E.~J., Knigge,
C., Bird, A.~J., Dean, A.~J., Clark, D.~J., Hill, A.~B., Molina, M., \&
Sguera, V.\ 2006, \mnras, 372, 234

\bibitem[\protect\citeauthoryear{Bassani et al.}{2005}]{basetal05}
Bassani, L., De Rosa, A., Bazzano, A., Bird, A.J., Dean, A.J.,
Gehrels, N., et al., 2005, ApJ, 634, L21

\bibitem[\protect\citeauthoryear{Bassani et al.}{2006}]{basetal06}
Bassani, L., Molina, M., Malizia, A., et al., 2006, ApJ, 636, L65

\bibitem[\protect\citeauthoryear{Beckmann et
al.}{2006}]{beckmann06} Beckmann V., Soldi S., Shrader C.~R.,
Gehrels N., Produit N., 2006, ApJ, 652, 126

\bibitem[\protect\citeauthoryear{B{\'e}langer et al.}{2006}]{belanger05}
B{\'e}langer G., Goldwurm, A., Renaud, M., et al.\ 2006, ApJ, 636, 275

\bibitem[\protect\citeauthoryear{Bikmaev et al.}{2006a}]{biketal06a}
Bikmaev, I.F., Sunyaev, R.A., Revnivtsev, M.G., \& Burenin, R.A.,
2006a, Astron. Lett., 32, 221

\bibitem[\protect\citeauthoryear{Bikmaev et al.}{2006b}]{biketal06b}
Bikmaev, I.F., Sunyaev, R.A., Revnivtsev, M.G., \& Burenin, R.A.,
2006b, Astron. Lett., 32, 588

\bibitem[Bird et al.(2006a)]{biretal06a} Bird, A.~J., Barlow, E.~J., Bassani, L.,
et al.\ 2006, \apj, 636, 765

\bibitem[\protect\citeauthoryear{Bird et al.}{2006b}]{biretal06b} Bird A. J.,Malizia A., Bazzano A., Barlow E. J., Bassani L., Hill A. B., Belanger G., Capitanio F., et al.  2006, astro, arXiv:astro-ph/0611493

\bibitem[Bodaghee et al.(2006)]{boda2006} Bodaghee, A., Walter, R., Zurita Heras, J.,
et al.\ 2006, \aap, 447, 1027

\bibitem[Brandt et al.(2005)]{brandt05} Brandt, S., Kuulkers, E., Bazzano, A., et al.\
2005, Astron. Telegram, 622, 1

\bibitem[Brandt et al.(2006)]{brandt06} Brandt, S., Budtz-J{\o}rgensen, C.,
\& Chenevez, J.\ 2006, Astron. Telegram, 778, 1

\bibitem[Burenin et al.(2006a)]{bureetal06a} Burenin, R.,
Mescheryakov, A., Revnivtsev, M., Bikmaev, I., \& Sunyaev, R.\ 2006a, The
Astronomer's Telegram, 880, 1

\bibitem[Burenin et al.(2006b)]{bureetal06b} Burenin, R.,
Mescheryakov, A., Sazonov, S., Revnivtsev, M., Bikmaev, I., \& Sunyaev, R.\
2006b, Astron. Telegram, 883, 1

\bibitem[Bykov et al.(2004)]{bykov2004} Bykov, A., Krassilshchikov, A., Uvarov, Yu.,
et al.\ 2004, \aap, 427, L21

\bibitem[Bykov et al.(2006)]{bykov2006} Bykov, A., Krassilshchikov, A., Uvarov, Yu.,
et al.\ 2006, \apjl, 649, L21

\bibitem[Chelovekov et al.(2006)]{cheletal06} Chelovekov, I., Grebenev, S., Sunyaev, R.\
2006, Astronomy Letters, 32, 456

\bibitem[Chernyakova et al.(2003)]{cheretal03} Chernyakova, M.,
Lutovinov, A., Capitanio, F., Lund, N., \& Gehrels, N.\ 2003, Astron. Telegram, 157, 1

\bibitem[Chernyakova et al.(2005a)]{cheetal05a} Chernyakova, M., Courvoisier, T.,
Rodriguez, J., Lutovinov, A.\ 2005a, Astron. Telegram, 519, 1

\bibitem[Chernyakova et al.(2005b)]{cheetal05b} Chernyakova, M.,
Lutovinov, A., Rodr{\'{\i}}guez, J., \& Revnivtsev, M.\ 2005b, \mnras, 364,
455

\bibitem[Cornelisse et al.(2006)]{cornalisse06} Cornelisse, R.,
Charles, P.~A., \& Robertson, C.\ 2006, \mnras, 366, 918

\bibitem[Courvoisier et al.(2003)]{courv2003} Courvoisier, T., Walter, R.,
Rodriguez, J., Bouchet, L., Lutovinov A.\ 2003, IAUC 8063, 1

\bibitem[Crawford et al.(1970)]{crawford1970} Crawford, D.~F.,
Jauncey, D.~L., \& Murdoch, H.~S.\ 1970, \apj, 162, 405

\bibitem[\protect\citeauthoryear{Donato et al.}{2005}]{donetal05}
Donato, D., Sambruna, R.M., \& Gliozzi, M., 2005, A\&A, 433, 1163

\bibitem[\protect\citeauthoryear{Davis \&
Peebles}{1983}]{davis83} Davis M., Peebles P.~J.~E., 1983, ARA\&A,
21, 109

\bibitem[den Hartog et al.(2004a)]{denhar2004a} den Hartog, P., Hermsen, W., Kuiper, L.,
et al.\ 2004a, Astron. Telegram, 261, 1

\bibitem[den Hartog et al.(2004b)]{denhar2004b} den Hartog, P., Kuiper, L., Corbet, R.,
et al.\ 2004b, Astron. Telegram, 281, 1

\bibitem[den Hartog et al.(2005)]{denhar2005} den Hartog, P., Kuiper, L., Hermsen, W.,
et al.\ 2005, Astron. Telegram, 394, 1

\bibitem[Hopkins et al.(2006)]{hopkins06} Hopkins, P.~F.,
Hernquist, L., Cox, T.~J., Di Matteo, T., Robertson, B., \& Springel, V.\
2006, \apjs, 163, 1

\bibitem[Fenimore et al.(1981)]{fenimor81} Fenimore, E.~E., Cannon T.~M., \
  1981 Applied Optics, 20, 1858.

\bibitem[Forman et al.(1978)]{uhuru} Forman, W., Jones, C., 
Cominsky, L., Julien, P., Murray, S., Peters, G., Tananbaum, H., \& 
Giacconi, R.\ 1978, \apjs, 38, 357 

\bibitem[Fouqu{\'e} et al.(2001)]{fouque01} Fouqu{\'e}, P.,
Solanes, J.~M., Sanchis, T., \& Balkowski, C.\ 2001, \aap, 375, 770

\bibitem[Gaensicke et al.(2005)]{gaenetal05} Gaensicke, B.~T., Marsh, T.~R, Edge, A., et
al.\ 2005, Astron. Telegram, 463, 1

\bibitem[G{\'o}rski et al.(2005)]{healpix} G{\'o}rski, K.~M.,
Hivon, E., Banday, A.~J., Wandelt, B.~D., Hansen, F.~K., Reinecke, M., \&
Bartelmann, M.\ 2005, \apj, 622, 759

\bibitem[Gotz et al.(2006)]{gotz2006} Gotz, D., Schanne, S.,
Rodriguez, J., Leyder, J.-C., von Kienlin, A., Mowlavi, N., \& Mereghetti,
S.\ 2006, Astron. Telegram, 813, 1

\bibitem[Grebenev et al.(2004a)]{grebetal04a} Grebenev, S.~A.,
Ubertini, P., Chenevez, J., Orr, A., \& Sunyaev, R.~A.\ 2004a, Astron. Telegram, 275, 1

\bibitem[Grebenev et al.(2004b)]{grebetal04b} Grebenev, S., Ubertini, P., Chenevez, J.,
et al.\ 2004b, Astron. Telegram, 350, 1

\bibitem[Grebenev et al.(2005a)]{grebetal05a} Grebenev, S.~A.,
Molkov, S.~V., \& Sunyaev, R.~A.\ 2005a, Astron. Telegram, 467, 1

\bibitem[Grebenev et al.(2005b)]{grebetal05b} Grebenev, S.~A.,
Molkov, S.~V., \& Sunyaev, R.~A.\ 2005b, The Astronomer's Telegram, 616, 1

\bibitem[Grimm et al.(2002)]{grimm02} Grimm, H.-J., Gilfanov,
M., \& Sunyaev, R.\ 2002, \aap, 391, 923

\bibitem[Gros et al.(2003)]{gros2003} Gros, A., Goldwurm, A.,
Cadolle-Bel, M., Goldoni, P., Rodriguez, J., Foschini, L., Del Santo, M.,
\& Blay, P.\ 2003, \aap, 411, L179

\bibitem[Halpern(2006)]{halp2006} Halpern, J.~P.\ 2006, Astron. Telegram, 847, 1

\bibitem[Hannikainen et al.(2003)]{hannetal03} Hannikainen, D., Rodriguez, J., Pottschmidt, K.
et al.\ 2003, IAUC 8088, 1

\bibitem[Hanski et al.(2001)]{hanski01} Hanski, M.~O., Theureau,
G., Ekholm, T., \& Teerikorpi, P.\ 2001, \aap, 378, 345

\bibitem[in't Zand \& Heise (2004)]{intzand04} in't Zand, J.~J.~M., \& Heise, J.\
2004, Astron. Telegram, 362, 1

\bibitem[in't Zand (2005)]{intzand05} in't Zand, J.~J.~M. \ 2005, \aap, 441, L1

\bibitem[Karasev et al.(2007)]{karasev07} Karasev, D., Lutovinov, A., Grebenev, S.\
2007, Astronomy Letters, 33, 135

\bibitem[Keek et al.(2006)]{keeketal06} Keek, S., Kuiper, L., \&
Hermsen, W.\ 2006, The Astronomer's Telegram, 810, 1

\bibitem[\protect\citeauthoryear{Kocevski \&
Ebeling}{2006}]{kocetal06} Kocevski D.~D., Ebeling H., 2006, ApJ,
645, 1043

\bibitem[Kormendy(2001)]{kormendy2001} Kormendy, J.\ 2001, Revista 
Mexicana de Astronomia y Astrofisica Conference Series, 10, 69 

\bibitem[Kormendy \& Richstone(1995)]{kormendy1995} Kormendy, J., \&
Richstone, D.\ 1995, \araa, 33, 581

\bibitem[Koyama et al.(1995)]{koyama1995} Koyama, K., Petre, R., Gotthelf,
E.~V., Hwang, U., Matsuura, M., Ozaki, M., \& Holt, S.~S.\ 1995, \nat, 378,
255

\bibitem[Koyama et al.(1997)]{koyama1997} Koyama, K., Kinugasa, K.,
Matsuzaki, K., Nishiuchi, M., Sugizaki, M., Torii, K., Yamauchi, S., \&
Aschenbach, B.\ 1997, \pasj, 49, L7

\bibitem[Kretschmar et al.(2004)]{kretetal04} Kretschmar, P., Mereghetti, S.,
Hermsen, W., et al.\ 2004, Astron. Telegram, 345, 1

\bibitem[\protect\citeauthoryear{Krivonos et al.}{2005a}]{krietal05}
Krivonos, R., Vikhlinin, A., Churazov, E., Lutovinov, A., Molkov, S.,
\& Sunyaev, R., 2005a, ApJ, 625, 89

\bibitem[Krivonos et al.(2005b)]{krivetal05b} Krivonos, R., Molkov,
S., Revnivtsev, M., Grebenev, S., Sunyaev, R., \& Paizis, A.\ 2005b, Astron.
Telegram, 545, 1

\bibitem[Krivonos et al.(2006)]{krietal06a} Krivonos, R., 
Revnivtsev, M., Sazonov, S., Churazov, E., \& Sunyaev, R.\ 2006, IAU 
Symposium, 230, 455 

\bibitem[Krivonos et al.(2006)]{krivetal06} Krivonos, R., Revnivtsev, M., Churazov, E.,
et al.\ 2006, \aap, accepted (astro-ph/0605420)

\bibitem[Kuiper et al.(2006)]{kuietal06} Kuiper, L., Keek, S.,
Hermsen, W., Jonker, P.~G., \& Steeghs, D.\ 2006, Astron.
Telegram, 684, 1

\bibitem[Kuulkers et al.(2003)]{kuulk2003} Kuulkers, E., Lutovinov, A., Parmar, A.,
et al.\ 2003, Astron. Telegram, 149, 1

\bibitem[Kuulkers et al.(2006)]{kuulk2006} Kuulkers, E., Shaw, S., Paizis, A.,
et al.\ 2006, Astron. Telegram, 874, 1

\bibitem[Lahav et al.(1988)]{lahav1988} Lahav, O., Lynden-Bell, D., \&
Rowan-Robinson, M.\ 1988, \mnras, 234, 677

\bibitem[Lahav et al.(1989)]{lahav1989} Lahav, O., Fabian, A.~C.,
Edge, A.~C., \& Putney, A.\ 1989, \mnras, 238, 881

\bibitem[Levine et al.(1984)]{levine84} Levine, A.~M., et al.\ 
1984, \apjs, 54, 581 

\bibitem[Liu et al.(2000)]{liuetal00} Liu, Q.~Z., van Paradijs,
J., \& van den Heuvel, E.~P.~J.\ 2000, \aaps, 147, 25

\bibitem[Liu et al.(2001)]{liuetal01} Liu, Q.~Z., van Paradijs,
J., \& van den Heuvel, E.~P.~J.\ 2001, \aap, 368, 1021

\bibitem[Lubinski et al.(2005)]{lubietal05} Lubinski, P., Gadolle Bel, M., von Kienlin, A.,
et al.\ 2005, Astron. Telegram, 469, 1

\bibitem[Lutovinov \& Revnivtsev(2003)]{lutrev03} Lutovinov,
A.~A., \& Revnivtsev, M.~G.\ 2003, Astronomy Letters, 29, 719

\bibitem[Lutovinov et al.(2003a)]{lutetal03a}  Lutovinov, A., Walter, R., Belanger, R.,
et al.\ 2003a, Astron. Telegram, 155, 1

\bibitem[Lutovinov et al.(2003b)]{lutetal03b} Lutovinov, A., Shaw, S., Foschini, L.,
Paul, J.\ 2003b, Astron. Telegram, 154, 1

\bibitem[Lutovinov et al.(2003c)]{lutetal03c} Lutovinov, A., Rodriguez, J., Produit, N.,
Paul, J.\ 2003c, Astron. Telegram, 151, 1

\bibitem[Lutovinov et al.(2004a)]{lutetal04a} Lutovinov, A., Rodriguez, J., Budtz-Jorgensen, C.,
et al.\ 2004a, Astron. Telegram, 329, 1

\bibitem[Lutovinov et al.(2004b)]{lutetal04b} Lutovinov, A., Bel,
M.~C., Belanger, G., Goldwurm, A., Budtz-Jorgensen, C., Mowlavi, N., Paul,
J., \& Orr, A.\ 2004, Astron. Telegram, 328, 1

\bibitem[Lutovinov et al.(2005a)]{lutetal05a} Lutovinov, A., Rodriguez, J.,
Revnivtsev, M., \& Shtykovskiy, P.\ 2005a, \aap, 433, L41

\bibitem[Lutovinov et al.(2005b)]{lutetal05b} Lutovinov, A.,
Revnivtsev, M., Gilfanov, M., et al.\ 2005b, \aap, 444, 821

\bibitem[Lutovinov et al.(2005c)]{lutetal05c} Lutovinov, A.,
Revnivtsev, M., Molkov, S., Sunyaev, R.\ 2005c, \aap, 430,997

\bibitem[Lynden-Bell et al.(1988)]{lynden1988} Lynden-Bell, D.,
Faber, S.~M., Burstein, D., Davies, R.~L., Dressler, A., Terlevich, R.~J.,
\& Wegner, G.\ 1988, \apj, 326, 19

\bibitem[Malizia et al.(2005)]{maletal05} Malizia, A., Bassani, L.,
Stephen, J., et al.\ 2005, \apjl, 630, L157

\bibitem[\protect\citeauthoryear{Masetti et al.}{2004}]{masetal04}
Masetti, N., Palazzi, E., Bassani, L., Malizia, A., \& Stephen, J.B.,
2004, A\&A, 426, L41

\bibitem[\protect\citeauthoryear{Masetti et al.}{2005}]{masetal05}
Masetti, N., Bassani, L., Bird, A., Bassano, A.\ 2005, Astron.
Telegram, 528,1

\bibitem[Masetti et al.(2006a)]{masetal06a} Masetti, N., Morelli,
L., Palazzi, E., Stephen, J., Bazzano, A., Dean, A.~J., Walter, R., \&
Minniti, D.\ 2006a, Astron. Telegram, 783, 1

\bibitem[\protect\citeauthoryear{Masetti et al.}{2006b}]{masetal06b}
Masetti, N., Morelli, L., Palazzi, E., Galaz, G., Bassani, L.,
Bazzano, A., et al., 2006b, \aap, 459, 21

\bibitem[\protect\citeauthoryear{Masetti et al.}{2006c}]{masetal06c}
Masetti, N., Pretorius, M.L., Palazzi, E., Bassani, L., Bazzano, A.,
\& Bird, A.J, 2006c, A\&A, 449, 1139

\bibitem[Masetti et al.(2006)]{masetal06d} Masetti, N., Bassani,
L., Dean, A.~J., Ubertini, P., \& Walter, R.\ 2006d, The Astronomer's
Telegram, 715, 1

\bibitem[\protect\citeauthoryear{Masetti et al.}{2006e}]{masetal06e}
Masetti, N., Bassani, L., Bazzano, A., Bird, A.J., Dean, A.J.,
Malizia, A., et al., 2006e, A\&A, 455, 11

\bibitem[Miyaji \& Boldt(1990)]{miyaji1990} Miyaji, T., \& Boldt, E.\ 1990,
\apjl, 353, L3

\bibitem[Molkov et al.(2003)]{molkov2003} Molkov, S., Mowlavi, N., Goldwurm, A.,
et al.\ 2003, Astron. Telegram, 176, 1

\bibitem[\protect\citeauthoryear{Molkov et al.}{2004}]{molkov2004}
Molkov S.~V., Cherepashchuk A.~M., Lutovinov A.~A., Revnivtsev M.~G.,
Postnov K.~A., Sunyaev R.~A., 2004, AstL, 30, 534

\bibitem[\protect\citeauthoryear{Morelli et al.}{2006}]{moretal06}
Morelli, L., Masetti, N., Bassani, L., et al., 2006, Astron. Telegram, 785, 1

\bibitem[Negueruela et al.(2005)]{neguetal05} Negueruela, I., Smith, D., Chaty, S. \
2005, Astron. Telegram,  470, 1

\bibitem[Negueruela \& Smith (2006)]{negsmi06} Negueruela, I., \& Smith, D.\
2006, Astron. Telegram,  831, 1

\bibitem[\protect\citeauthoryear{Neronov et
al.}{2005}]{neronov05} Neronov A., Chernyakova M., Courvoisier
T.~J.~-., Walter R., 2005, astro, arXiv:astro-ph/0506437

\bibitem[Palmer et al.(2005)]{palmetal05} Palmer, D.~M.,
Barthelmey, S.~D., Cummings, J.~R., Gehrels, N., Krimm, H.~A., Markwardt,
C.~B., Sakamoto, T., \& Tueller, J.\ 2005, The Astronomer's Telegram, 546,
1
\bibitem[Patel et al.(2004)]{patel2004} Patel, S., Kouveliotou, C., Tennant, A.,
et al.\ 2004, \apj, 602, L45

\bibitem[Pavlinskii et al.(1992)]{pavetal92} Pavlinskii, M.~N.,
Grebenev, S.~A., \& Syunyaev, R.~A.\ 1992, Soviet Astronomy Letters,
18, 88


\bibitem[Pfeffermann \& Aschenbach(1996)]{pfeffermann1996} Pfeffermann,
E., \& Aschenbach, B.\ 1996, Roentgenstrahlung from the Universe, 267

\bibitem[Plionis \& Kolokotronis(1998)]{plionis1998} Plionis, M., \&
Kolokotronis, V.\ 1998, \apj, 500, 1

\bibitem[Produit et al.(2003)]{prodetal03} Produit, N., Ballet, J., Mowlavi,
N.\ 2003, Astron. Telegram., 278, 1

\bibitem[Protheroe et al.(1980)]{protheroe1980} Protheroe, R.~J.,
Wolfendale, A.~W., \& Wdowczyk, J.\ 1980, \mnras, 192, 445

\bibitem[Rea et al.(2006)]{reaetal06} Rea, N., Testa, V., Israel, G., et al.\
2006, Astron. Telegram, 713, 1

\bibitem[\protect\citeauthoryear{Rees}{1980}]{rees80} Rees
M.~J., 1980, IAUS, 92, 207

\bibitem[Reig et al.(2005)]{reig2005} Reig, P., Negueruela, I.,
Papamastorakis, G., Manousakis, A., \& Kougentakis, T.\ 2005, \aap, 440,
637

\bibitem[Renaud et al.(2006a)]{renaud2006a} Renaud, M., Gros, A.,
Lebrun, F., Terrier, R., Goldwurm, A., Reynolds, S., \& Kalemci, E.\ 2006a,
\aap, 456, 389

\bibitem[Renaud et al.(2006b)]{renaud2006b} Renaud, M.,
B{\'e}langer, G., Paul, J., Lebrun, F., \& Terrier, R.\ 2006b, \aap, 453, L5

\bibitem[Revnivtsev et al.(2003a)]{revetal03a} Revnivtsev, M.~G.,
Sazonov, S.~Y., Gilfanov M.~R., Sunyaev, R.~A.\ 2003a, Astronomy Letters, 29, 587

\bibitem[Revnivtsev et al.(2003b)]{revetal03b} Revnivtsev, M., Tuerler, M., Del Santo, M.,
et al.\ 2003b, IAUC 8097, 1

\bibitem[Revnivtsev et al.(2003c)]{revetal03c} Revnivtsev, M., Chernyakova, M., Capitanio, F.,
et al.\ 2003c, Astron. Telegram, 132, 1

\bibitem[\protect\citeauthoryear{Revnivtsev et al.}{2004a}]{revetal04}
Revnivtsev, M., Sazonov, S., Jahoda, K., \& Gilfanov, M., 2004a,
A\&A, 418, 927

\bibitem[Revnivtsev et al.(2004)]{revetal04sigma} Revnivtsev, M.~G., 
Sunyaev, R.~A., Gilfanov, M.~R., Churazov, E.~M., Goldwurm, A., Paul, J., 
Mandrou, P., \& Roques, J.~P.\ 2004, Astronomy Letters, 30, 527 

\bibitem[Revnivtsev et al.(2004b)]{revetal04B2} Revnivtsev, M., Sazonov, S., Churazov, E.,
et al.\ 2004b, \aap, 425, L49

\bibitem[Revnivtsev et al.(2004c)]{revetal04c} Revnivtsev, M., Sunyaev, R., Varshalovich, D.,
et al.\ 2004c, Astronomy Letters, 30, 382

\bibitem[Revnivtsev et al.(2006)]{revetal06} Revnivtsev, M.~G.,
Sazonov, S.~Y., Molkov, S.~V., Lutovinov, A.~A., Churazov, E.~M., \&
Sunyaev, R.~A.\ 2006, Astronomy Letters, 32, 145

\bibitem[Richstone et al.(1998)]{richstone1998} Richstone, D., et
al.\ 1998, \nat, 395, A14

\bibitem[Rodriguez et al.(2004)]{rodetal04} Rodriguez, J., Domingo Garau, A.,
Grebenev, S., et al.\ 2004, Astron. telegram, 340, 1

\bibitem[Rodriguez et al.(2005)]{rodetal05} Rodriguez, J.,
Cabanac, C., Hannikainen, D.~C., Beckmann, V., Shaw, S.~E., \& Schultz, J.\
2005, \aap, 432, 235

\bibitem[\protect\citeauthoryear{Rowan-Robinson et
al.}{2000}]{rowanetal00} Rowan-Robinson M., et al., 2000, MNRAS,
314, 375

\bibitem[Saunders et al.(2000)]{saunders00} Saunders, W., et al.\
2000, \mnras, 317, 55

\bibitem[\protect\citeauthoryear{Sazonov et al.}{2005}]{sazetal05}
Sazonov, S., Churazov, E., Revnivtsev, M., Vikhlinin, A., \& Sunyaev,
R., 2005, A\&A, 444, L37

\bibitem[\protect\citeauthoryear{Sazonov et al.}{2007}]{sazetal07}
Sazonov, S.,  Revnivtsev, M., Krivonos, R., Churazov, \& Sunyaev,
R., 2006, A\&A, 462, 57

\bibitem[Sguera et al.(2006)]{sguetal06} Sguera, V., Bazzano, A., Bird, A., et al.\
2006, \apj, 646, 452

\bibitem[Skinner et al.(1981)]{skinner87a} Skinner, G.~K. et al. 1987, \
  \apss,  136, 337-349.


\bibitem[Smith et al.(2006)]{smietal06} Smith, D.~M., Heindl,
W.~A., Markwardt, C.~B., Swank, J.~H., Negueruela, I., Harrison, T.~E., \&
Huss, L.\ 2006, \apj, 638, 974

\bibitem[Slane et al.(1999)]{slane1999} Slane, P., Gaensler, B.~M., Dame,
T.~M., Hughes, J.~P., Plucinsky, P.~P., \& Green, A.\ 1999, \apj, 525, 357

\bibitem[Sunyaev et al.(2003a)]{sunyaev03a} Sunyaev, R., Lutovinov, A., Molkov, S., Deluit, S.\
2003a, Astron. Telegram, 181, 1

\bibitem[Sunyaev et al.(2003b)]{sunyaev03b} Sunyaev, R., Grebenev, S., Lutovinov, A.,
et al.\ 2003b, Astron. Telegram, 192, 1

\bibitem[Tonry et al.(2000)]{tonry00} Tonry, J.~L., Blakeslee,
J.~P., Ajhar, E.~A., \& Dressler, A.\ 2000, \apj, 530, 625


\bibitem[Tomsick et al.(2003)]{tomsick2003} Tomsick, J., Lingenfelter, R., Walter, R.,
et al.\ 2003, IAUC 8076, 1

\bibitem[Tomsick et al.(2004)]{tomsick2004} Tomsick, J., Lingenfelter, R., Corbel, S.,
Goldwurm, A., Kaaret, P.\ 2004, Astron. Telegram, 224, 1

\bibitem[Tomsick et al.(2006)]{tomsick2006} Tomsick, J., Chaty, S.,
Rodriguez, J., et al.\ 2006, ApJ, 647, 1309

\bibitem[Toor \& Seward(1974)]{toor74} Toor, A., \& Seward,
F.~D.\ 1974, \aj, 79, 995

\bibitem[Torres et al.(2005)]{torretal05} Torres, M.~A.~P., et
al.\ 2005, Astron. Telegram, 551, 1

\bibitem[Tueller et al.(2005)]{tueletal05} Tueller, J., Barthelmy, S., Burrows, D., et al.\
2005, Astron. Telegram, 669, 1

\bibitem[Turler et al.(2005)]{turletal05} Turler, M., Bel, M.~C.,
Diehl, R., Westergaard, N.-J., McBreen, B., Williams, O.~R., Grebenev,
S.~A., \& Lutovinov, A.\ 2005, The Astronomer's Telegram, 624, 1

\bibitem[\protect\citeauthoryear{Ubertini et
al.}{2003}]{ibis} Ubertini P., et al., 2003, A\&A, 411, L131


\bibitem[Ubertini et al.(2005)]{ubeetal05} Ubertini, P., Bassani, L., Malizia, A., et al.\
2005, \apjl, 629, L109

\bibitem[Walter et al.(2003)]{walter2004} Walter, R., Bodaghee, A., Barlow, E.,
et al.\ 2003, Astron. Telegram, 229, 1

\bibitem[Walter et al.(2006)]{waletal06} Walter, R., Zurita Heras, J., Bassani, L.,
et al.\ 2006, \aap, 453, 133

\bibitem[\protect\citeauthoryear{Winkler et al.}{2003}]{integral} Winkler, C.,
Courvoisier, T., Di Cocco, G., et al.\ 2003, A\&A, 411, L1

\end{thebibliography}

\clearpage

\tabletypesize{\scriptsize} 
\begin{deluxetable}{rlrrrclc} 
\tablecaption{INTEGRAL Catalog \label{tab:catalog}} 
\tablewidth{0pt} 
\tablehead{ 
\colhead{Id} & 
\colhead{Name} & 
\colhead{RA} & 
\colhead{Dec} & 
\colhead{$F_{\rm 17-60~keV}$} & 
\colhead{Type} & 
\colhead{Counterpart} & 
\colhead{Notes} 
} 
\startdata 
1 & IGR J00234+6141  & 5.723 & 61.700 & $0.38\pm0.10$ &  CV  &    &  43,50 \\
2 & TYCHO SNR  & 6.334 & 64.150 & $0.64\pm0.10$ &  SNR   &    &   \\
3 & V709 Cas  & 7.205 & 59.300 & $3.91\pm0.11$ &  CV   &    &  36 \\
4 & IGR J00291+5934  & 7.254 & 59.563 & $4.01\pm0.11$ &  ~~~~~LMXB~~~~~  &    &   \\
5 & 87GB003300.9+593328  & 8.977 & 59.827 & $0.72\pm0.11$ &  AGN   &    &  1 \\
6 & IGR J00370+6122  & 9.286 & 61.386 & $0.60\pm0.11$ &  HMXB   &    &  45 \\
7 & MRK 348  & 12.181 & 31.947 & $5.18\pm0.58$ &  AGN   &  NGC 262  &   \\
8 & 1WGA J0053.8-722  & 13.526 & -72.468 & $2.36\pm0.38$ &  HMXB   &    &   \\
9 & Gamma Cas  & 14.176 & 60.712 & $3.17\pm0.12$ &  Star   &    &   \\
10 & SMC X-1  & 19.299 & -73.449 & $30.54\pm0.36$ &  HMXB  &    &  \\
11 & 1A 0114+650  & 19.516 & 65.289 & $7.64\pm0.14$ &  HMXB  &         &   \\
12 & 4U 0115+63	  & 19.625 & 63.746 & $2.01\pm0.14$ &  HMXB   &    &   \\
13 & NGC 0526A  & 20.951 & -34.925 & $3.84\pm0.81$ &  AGN   &    &   \\
14 & IGR J01363+6610  & 24.060 & 66.188 & $15.61\pm2.20$\tablenotemark{R185} &  HMXB  &    &  59,52 \\
15 & RX J0137.7+5814  & 24.443 & 58.221 & $0.77\pm0.20$ &    &    &   \\
16 & ESO 297- G 018  & 24.639 & -40.020 & $3.85\pm0.80$\tablenotemark{R374} &  AGN   &    &   \\
17 & 4U 0142+61  & 26.630 & 61.738 & $2.02\pm0.19$ &  AXP   &    &   \\
18 & RJ 0146.9+6121  & 26.744 & 61.351 & $2.57\pm0.20$ &  HMXB  &    &   \\
19 & IGR J01528-0326  & 28.208 & -3.450 & $1.14\pm0.21$ &  AGN   &  MCG -01-05-047  &  55 \\
20 & NGC 788  & 30.277 & -6.819 & $3.37\pm0.20$ &  AGN   &    &  \\
21 & IGR J02095+5226  & 32.392 & 52.458 & $2.76\pm0.50$ &  AGN   &  LEDA 138501 &   \\
 & & & & & & 1ES 0206+522  & \\
22 & MRK 1040  & 37.063 & 31.316 & $3.39\pm0.57$ &  AGN  &  NGC 931  &   \\
23 & IGR J02343+3229  & 38.599 & 32.475 & $2.71\pm0.44$ &  AGN   &  NGC 973 &  55 \\
 & & & & & & IRAS 02313+3217  & \\
24 & NGC 1052  & 40.267 & -8.236 & $1.48\pm0.30$ &  AGN  &    &   \\
25 & NGC 1068  & 40.687 & -0.010 & $1.32\pm0.21$ &  AGN  &    &   \\
26 & 4U 0241+61  & 41.262 & 62.464 & $3.32\pm0.42$ &  AGN  &    &  \\
27 & IGR J02466-4222  & 41.644 & -42.360 & $2.16\pm0.38$ &  AGN?   &  MCG -07-06-018  &  77,101 \\
28 & IGR J02524-0829  & 43.115 & -8.486 & $2.20\pm0.47$ &  AGN?  &  MCG-02-08-014   &   \\
29 & NGC 1142  & 43.804 & -0.186 & $3.22\pm0.25$ &  AGN   &  NGC 1144  &   \\
30 & PERSEUS CLUSTER\tablenotemark{a}  & 49.973 & 41.509 & $2.49\pm0.24$ &  Cluster   &    &   \\
31 & 1H 0323+342  & 51.140 & 34.168 & $1.91\pm0.33$ &  AGN   &    &   \\
32 & GK Per  & 52.777 & 43.880 & $1.82\pm0.27$ &  CV   &    &   \\
33 & IGR J03334+3718  & 53.362 & 37.313 & $1.37\pm0.29$ &  AGN   &    &  60,77 \\
34 & NGC 1365  & 53.428 & -36.170 & $2.30\pm0.46$ &  AGN   &    &  \\
35 & V 0332+53  & 53.751 & 53.172 & $135.14\pm0.44$ &  HMXB   &    &   \\
36 & 4U 0352+30   & 58.849 & 31.036 & $36.69\pm0.62$ &  HMXB  &  X Per  &   \\
37 & 3C111  & 64.581 & 38.013 & $5.47\pm0.61$ &  AGN   &    &   \\
38 & LEDA 168563  & 73.044 & 49.531 & $3.09\pm0.78$ &  AGN  &  1RXS J045205.0+493248  &   \\
39 & ESO 033-G002  & 74.001 & -75.538 & $1.36\pm0.19$ &  AGN   &    &   \\
40 & IGR J05007-7047  & 75.203 & -70.775 & $1.18\pm0.17$ &  HMXB  &  IGR J05009-7044  &  2 \\
41 & V1062 Tau  & 75.617 & 24.732 & $3.67\pm0.62$\tablenotemark{R102} &  CV  &    &   \\
42 & IRAS 05078+1626  & 77.705 & 16.513 & $4.14\pm0.53$ &  AGN   &    &   \\
43 & 4U 0513-40       & 78.534 & -40.069 & $3.39\pm0.60$ &  LMXB   &    &  \\
44 & AKN 120  & 79.026 & -0.140 & $5.86\pm2.46$ &  AGN   &    &   \\
45 & IGR J05305-6559\tablenotemark{b} & 82.636 & -65.984 & $1.72\pm0.34$ &    &    &   \\
46 & LMC X-4  & 83.210 & -66.367 & $15.95\pm0.17$ &  HMXB   &    &   \\
47 & Crab         & 83.632 & 22.016 & $1000.00\pm0.41$ &  PSR   &         &   \\
48 & TW Pic  & 83.689 & -57.988 & $0.97\pm0.27$ &  CV  &    &   \\
49 & A 0535+262  & 84.735 & 26.324 & $3.45\pm0.42$ &  HMXB   &    &   \\
50 & LMC X-1  & 84.912 & -69.748 & $3.72\pm0.17$ &  HMXB   &    &   \\
51 & PSR 0540-69  & 85.005 & -69.338 & $1.70\pm0.17$ &  PSR   &    &   \\
52 & BY Cam  & 85.713 & 60.868 & $2.48\pm0.51$ &  CV   &    &  36 \\
53 & MCG 8-11-11  & 88.801 & 46.437 & $3.92\pm1.05$ &  AGN   &    &  \\
54 & IRAS 05589+2828  & 90.601 & 28.461 & $3.38\pm0.71$ &  AGN   &    &   \\
55 & ESO 005- G 004  & 92.575 & -86.554 & $1.76\pm0.32$ &  AGN  &    &   \\
56 & MRK 3  & 93.908 & 71.036 & $4.77\pm0.21$ &  AGN    &    &   \\
57 & 4U 0614+091  & 94.282 & 9.139 & $20.81\pm0.67$ &  LMXB  &    &   \\
58 & IGR J06239-6052\tablenotemark{b}  & 95.936 & -60.974 & $1.16\pm0.22$ &    &    &   \\
59 & MRK 6  & 103.048 & 74.423 & $2.55\pm0.21$ &  AGN   &    &   \\
60 & IGR J07264-3553  & 111.595 & -35.900 & $1.89\pm0.42$ &  AGN   &  LEDA 096373  &   \\
61 & EXO 0748-676  & 117.146 & -67.754 & $19.32\pm0.35$ &  LMXB  &    &   \\
62 & IGR J07563-4137  & 119.055 & -41.638 & $0.86\pm0.17$ &  AGN &  IGR J07565-4139 &  44,2 \\
 & & & & & & 2MASX J07561963-4137420  & \\
63 & IGR J07597-3842  & 119.934 & -38.727 & $2.03\pm0.18$ &  AGN  &    &  13,8 \\
64 & ESO 209-G012  & 120.496 & -49.734 & $1.16\pm0.17$ &  AGN   &    &   \\
65 & IGR J08023-6954  & 120.762 & -69.924 & $3.70\pm0.96$ &    &    &  23 \\
66 & PG 0804+761  & 122.952 & 76.102 & $2.01\pm0.43$ &  AGN &    &  \\
67 & Vela pulsar  & 128.835 & -45.182 & $7.03\pm0.12$ &  PSR   &    &   \\
68 & 4U 0836-429  & 129.354 & -42.894 & $29.46\pm0.12$ &  LMXB  &    &   \\
69 & FAIRALL 1146  & 129.621 & -35.983 & $1.14\pm0.17$ &  AGN   &    &  \\
70 & IGR J08408-4503  & 130.218 & -45.056 & $0.33\pm0.12$ &  HMXB  &    &  75  \\
71 & S5 0836+71  & 130.340 & 70.902 & $2.43\pm0.23$ &  AGN   &    &  1 \\
72 & Vela X-1  & 135.531 & -40.555 & $187.16\pm0.14$ &  HMXB  &    &   \\
73 & IGR J09026-4812  & 135.648 & -48.221 & $1.32\pm0.13$ &   &   &  44   \\
74 & IRAS 09149-6206  & 139.043 & -62.330 & $1.44\pm0.18$ &  AGN  &    &   \\
75 & X 0918-548  & 140.102 & -55.196 & $3.34\pm0.15$ &  ~~~~~LMXB~~~~~   &    &   \\
76 & SWIFT J0920.8-0805  & 140.213 & -8.086 & $2.51\pm0.70$ &  AGN   &  MCG-01-24-012  &   \\
77 & IGR J09251+5219  & 141.274 & 52.331 & $4.09\pm0.79$ &  AGN   &  Mrk 110  &   \\
78 & IGR J09446-2636\tablenotemark{c}  & 146.124 & -26.628 & $2.73\pm0.52$ &  AGN  &  1RXS J094436.5-263353 &   \\
 & & & & & & 6dF J0944370-263356  & \\
79 & NGC 2992  & 146.431 & -14.335 & $3.55\pm0.27$ &  AGN   &    &   \\
80 & MCG-5-23-16  & 146.916 & -30.947 & $6.82\pm0.59$ &  AGN  &  ESO 434-G040  &   \\
81 & IGR J09522-6231  & 148.025 & -62.523 & $0.82\pm0.15$ &    &    &  77 \\
82 & NGC 3081  & 149.859 & -22.816 & $3.23\pm0.39$ &  AGN   &    &   \\
83 & IGR J10095-4248  & 152.449 & -42.800 & $1.50\pm0.27$ &  AGN  &  ESO 263-G013  &   \\
84 & GRO J1008-57  & 152.447 & -58.298 & $4.11\pm0.12$ &  HMXB   &    &   \\
85 & IGR J10100-5655  & 152.529 & -56.914 & $1.20\pm0.13$ &  HMXB   &    &  62,8 \\
86 & IGR J10109-5746  & 152.753 & -57.795 & $1.01\pm0.13$ &  SimbStar?   &  2RXP J101103.0-574810  &  23,67 \\
87 & NGC 3227  & 155.876 & 19.867 & $6.32\pm0.58$ &  AGN   &    &   \\
88 & IGR J10252-6829  & 156.287 & -68.460 & $3.14\pm0.87$ &     &    &  23 \\
89 & NGC 3281  & 157.935 & -34.855 & $2.70\pm0.45$ &  AGN   &    &   \\
90 & 3U 1022-55  & 159.401 & -56.801 & $4.11\pm0.47$\tablenotemark{R85} &  HMXB   &    &   \\
91 & IGR J10386-4947  & 159.676 & -49.789 & $1.03\pm0.17$ &  AGN  &  SWIFT J1038.8-4942  &  20 \\
92 & IGR J10404-4625  & 160.124 & -46.391 & $1.47\pm0.24$ &  AGN  &  LEDA 93974  &  44,4 \\
93 & $\eta$ Car  & 161.189 & -59.719 & $0.68\pm0.12$ &  Star   &    &   \\
94 & IGR J11085-5100  & 167.144 & -51.014 & $0.19\pm0.17$ &     &    &  23 \\
95 & Cen X-3  & 170.306 & -60.628 & $52.23\pm0.13$ &  HMXB  &    &   \\
96 & IGR J11215-5952  & 170.429 & -59.869 & $1.09\pm0.13$ &  HMXB   &    &  12,56  \\
97 & IGR J11305-6256  & 172.779 & -62.945 & $3.45\pm0.14$ &  XRB &    &  34,4 \\
98 & IGR J11321-5311  & 173.047 & -53.200 & $22.36\pm2.14$ &     &    &  58 \\
99 & NGC 3783  & 174.739 & -37.766 & $8.58\pm1.30$ &  AGN   &    &   \\
100 & IGR J11395-6520  & 174.858 & -65.406 & $10.68\pm0.86$\tablenotemark{R88} &  RS CVn?  &  HD 101379  &   \\
101 & IGR J11435-6109  & 176.031 & -61.106 & $0.88\pm0.15$ &  HMXB   &    &  14,18 \\
102 & A 1145.1-6141  & 176.870 & -61.956 & $22.91\pm0.15$ &  HMXB  &    &   \\
103 & 4U 1145-619\tablenotemark{b} & 177.000 & -62.207 & $3.57\pm0.15$ &  HMXB  &    &   \\
104 & IGR J12026-5349  & 180.686 & -53.823 & $1.70\pm0.21$ &  AGN  &  WKK0560  &  23,2 \\
105 & NGC 4051  & 180.781 & 44.525 & $2.11\pm0.48$ &  AGN   &    &   \\
106 & NGC 4138  & 182.352 & 43.672 & $1.67\pm0.34$ &  AGN   &    &   \\
107 & NGC 4151  & 182.634 & 39.408 & $33.11\pm0.30$ &  AGN    &    &   \\
108 & 1ES 1210-646     & 183.269 & -64.917 & $0.66\pm0.18$ &    &    &   \\
109 & IGR J12143+2933  & 183.597 & 29.561 & $0.67\pm0.25$ &  AGN   &  WAS 49B  &   \\
110 & NGC 4253  & 184.592 & 29.825 & $1.07\pm0.21$ &  AGN   &    &   \\
111 & NGC 4258  & 184.747 & 47.309 & $1.33\pm0.39$ &  AGN   &    &   \\
112 & PKS 1219+04  & 185.588 & 4.230 & $0.91\pm0.16$ &  AGN   &    &  \\
113 & MRK 50  & 185.860 & 2.676 & $0.91\pm0.16$ &  AGN   &    &   \\
114 & NGC 4388  & 186.444 & 12.664 & $12.50\pm0.21$ &  AGN   &    &   \\
115 & NGC 4395  & 186.462 & 33.565 & $1.08\pm0.20$ &  AGN   &    &   \\
116 & GX 301-2  & 186.651 & -62.772 & $122.97\pm0.19$ &  HMXB  &    &   \\
117 & XSS J12270-4859  & 186.978 & -48.907 & $1.39\pm0.33$ &  CV   &    &  63 \\
118 & 3C273  & 187.271 & 2.050 & $9.66\pm0.16$ &  AGN    &    &   \\
119 & IGR J12349-6434  & 188.724 & -64.565 & $3.56\pm0.19$ &  SimbStar?   &  RT Cru  &  17,68 \\
120 & NGC 4507  & 188.908 & -39.905 & $7.64\pm0.34$ &  AGN   &    &   \\
121 & IGR J12391-1612  & 189.792 & -16.186 & $2.10\pm0.43$ &  AGN   &  LEDA 170194 &  23,2 \\
 & & & & & & XSS 12389-1614  & \\
122 & NGC 4593  & 189.910 & -5.347 & $4.09\pm0.18$ &  AGN    &    &   \\
123 & WKK 1263  & 190.356 & -57.841 & $0.89\pm0.22$ &  AGN  &    &   \\
124 & PKS 1241-399  & 191.057 & -40.115 & $1.24\pm0.28$ &  AGN   &    &   \\
125 & 4U 1246-588  & 192.386 & -59.090 & $2.36\pm0.21$ &  HMXB?   &    &  40 \\
126 & 3C279  & 194.030 & -5.779 & $1.05\pm0.19$ &  AGN    &    &  \\
127 & 2S 1254-690  & 194.392 & -69.296 & $2.60\pm0.24$ &  LMXB  &    &  41 \\
128 & Coma  & 194.865 & 27.938 & $1.58\pm0.15$ &  Cluster   &    &   \\
129 & 1RXP J130159.6-635806\tablenotemark{b} & 195.495 & -63.969 & $1.64\pm0.20$ &  HMXB   &    &  51 \\
130 & PSR B1259-63\tablenotemark{b} & 195.699 & -63.836 & $1.01\pm0.19$ &  HMXB   &    &   \\
131 & MRK 783  & 195.741 & 16.361 & $0.86\pm0.22$ &  AGN   &    &   \\
132 & IGR J13038+5348  & 195.951 & 53.798 & $2.19\pm0.46$ &  AGN  &  MCG+09-21-096  &  60,77 \\
133 & NGC 4945  & 196.364 & -49.470 & $13.92\pm0.25$ &  AGN    &    &   \\
134 & ESO 323-G077  & 196.607 & -40.423 & $1.94\pm0.24$ &  AGN  &    &  \\
135 & IGR J13091+1137  & 197.270 & 11.619 & $2.44\pm0.29$ &  AGN   &  NGC 4992  &  23,2 \\
136 & IGR J13109-5552  & 197.689 & -55.865 & $1.29\pm0.21$ &    &  PMN J1310-5552  &  23 \\
137 & NGC 5033  & 198.350 & 36.572 & $0.84\pm0.19$ &  AGN   &    &   \\
138 & IGR J13149+4422\tablenotemark{c}  & 198.743 & 44.389 & $1.51\pm0.27$ &  AGN   &  Mrk 248  &   \\
139 & Cen A  & 201.363 & -43.019 & $39.19\pm0.22$ &  AGN  &    &   \\
140 & 4U 1323-619  & 201.643 & -62.136 & $5.28\pm0.18$ &  LMXB   &    &   \\
141 & IGR J13290-6323\tablenotemark{c}  & 202.268 & -63.392 & $2.30\pm0.37$\tablenotemark{R92} &    &    &    \\
142 & ESO 383-G018  & 203.332 & -34.030 & $1.22\pm0.27$ &  AGN   &    &   \\
143 & BH CVn  & 203.699 & 37.182 & $0.57\pm0.26$ &  RS CVn   &    &   \\
144 & MCG-6-30-15  & 203.990 & -34.288 & $2.53\pm0.26$ &  AGN   &  ESO 383-G035  &   \\
145 & NGC 5252  & 204.564 & 4.528 & $3.45\pm1.03$ &  AGN   &    &  \\
146 & MRK 268  & 205.420 & 30.395 & $1.22\pm0.21$ &  AGN   &    &   \\
147 & 4U 1344-60  & 206.894 & -60.615 & $4.16\pm0.17$ &  AGN   &    &  \\
148 & IC 4329A  & 207.333 & -30.309 & $11.28\pm0.36$ &  AGN    &    &   \\
149 & IGR J14003-6326  & 210.204 & -63.414 & $0.89\pm0.15$ &    &    &  57 \\
150 & V834 Cen  & 212.196 & -45.382 & $4.11\pm0.91$ &   CV    &   &   \\
151 & Circinus galaxy  & 213.290 & -65.342 & $13.11\pm0.18$ &  AGN    &    &   \\
152 & NGC 5506  & 213.312 & -3.220 & $9.32\pm0.46$ &  AGN   &    &   \\
153 & IGR J14175-4641  & 214.296 & -46.671 & $0.91\pm0.19$ &  AGN  &    &  23,8 \\
154 & NGC 5548  & 214.541 & 25.155 & $1.44\pm0.35$ &  AGN   &    &   \\
155 & ESO 511-G030  & 214.885 & -26.633 & $2.32\pm0.63$ &  AGN  &    &   \\
156 & IGR J14298-6715  & 217.388 & -67.251 & $0.89\pm0.18$ &    &    &  57 \\
157 & IGR J14331-6112  & 218.273 & -61.221 & $1.00\pm0.15$ &    &    &  57 \\
158 & IGR J14471-6414  & 221.528 & -64.284 & $0.86\pm0.15$ &    &    &  57 \\
159 & IGR J14471-6319  & 221.834 & -63.289 & $0.66\pm0.16$ &  AGN  &    &  23,8 \\
160 & IGR J14493-5534  & 222.311 & -55.589 & $1.15\pm0.16$ &  AGN   &  2MASX J14491283-5536194  &  22,9 \\
161 & IGR J14515-5542  & 222.887 & -55.691 & $1.08\pm0.15$ &  AGN   &  WKK 4374  &  62,8 \\
162 & IGR J14536-5522  & 223.421 & -55.363 & $1.20\pm0.16$ &  CV   &    &  62,63  \\
163 & IGR J14552-5133  & 223.846 & -51.571 & $0.98\pm0.16$ &  AGN   &  WKK 4438  &  23,8 \\
164 & IGR J14561-3738\tablenotemark{c}  & 224.055 & -37.632 & $0.98\pm0.18$ &  AGN  &  ESO 386- G 034  &  101 \\
165 & IC 4518A  & 224.427 & -43.125 & $1.70\pm0.16$ &  AGN   &    &   \\
166 & IGR J15094-6649  & 227.382 & -66.816 & $1.05\pm0.21$ &  CV   &    &  23,63 \\
167 & PSR 1509-58  & 228.480 & -59.145 & $8.85\pm0.15$ &  PSR   &    &   \\
168 & 4U 1516-569  & 230.167 & -57.168 & $7.64\pm0.15$ &  LMXB   &    &   \\
169 & IGR J15360-5750  & 234.014 & -57.806 & $0.93\pm0.15$ &    &    &  23 \\
170 & 4U 1538-522  & 235.600 & -52.385 & $16.42\pm0.13$ &  HMXB   &    &   \\
171 & XTE J1543-568  & 236.011 & -56.748 & $0.84\pm0.14$ &  HMXB   &    &   \\
172 & 4U 1543-624  & 236.964 & -62.578 & $2.27\pm0.18$ &  LMXB   &    &  \\
173 & NY Lup  & 237.052 & -45.472 & $4.10\pm0.14$ &  CV   &  1RXS J154814.5-452845  &   \\
174 & XTE J1550-564  & 237.751 & -56.474 & $28.65\pm0.14$ &  LMXB   &    &   \\
175 & IGR J15539-6142  & 238.468 & -61.676 & $0.85\pm0.17$ &    &    &  57 \\
176 & 4U 1556-605  & 240.363 & -60.716 & $0.84\pm0.16$ &  LMXB   &    &   \\
177 & WKK 6092  & 242.981 & -60.637 & $1.16\pm0.17$ &  AGN   &    &   \\
178 & 4U 1608-522  & 243.177 & -52.425 & $6.48\pm0.13$ &  LMXB   &    &   \\
179 & IGR J16167-4957  & 244.162 & -49.975 & $1.45\pm0.14$ &  CV   &  1RXS J161637.2-495847  &  86,36 \\
180 & IGR J16175-5059  & 244.357 & -50.972 & $0.60\pm0.13$ &  PSR &  PSR J1617-5055  &   \\
181 & IGR J16185-5928  & 244.635 & -59.468 & $1.21\pm0.16$ &  AGN  &  WKK 6471  &  23,8 \\
182 & IGR J16195-4945  & 244.893 & -49.755 & $1.80\pm0.14$ &  HMXB   &  AX J161929-4945  &  86,83 \\
183 & IGR J16195-2807  & 244.871 & -28.151 & $2.07\pm0.31$ &  RS CVn?   &  1RXS J161933.6-280736  &  44 \\
184 & Sco X-1  & 244.981 & -15.637 & $581.08\pm0.44$ &  LMXB  &    &   \\
185 & IGR J16207-5129  & 245.194 & -51.505 & $3.01\pm0.14$ &  HMXB   &    &  86,83 \\
186 & SWIFT J1626.6-5156  & 246.659 & -51.938 & $23.58\pm1.88$\tablenotemark{R399} &  ~~~~~LMXB~~~~~   &    &  78 \\
187 & 4U 1624-49  & 247.002 & -49.209 & $3.88\pm0.14$ &  LMXB   &    &   \\
188 & IGR J16318-4848  & 247.953 & -48.819 & $19.53\pm0.14$ &  HMXB  &    &  84,85 \\
189 & IGR J16320-4751  & 248.013 & -47.876 & $15.34\pm0.14$ &  HMXB  &  AX J1631.9-4752  &  87,25 \\
190 & 4U 1626-67  & 248.076 & -67.466 & $13.58\pm0.34$ &  LMXB  &    &   \\
191 & 4U 1630-47  & 248.503 & -47.391 & $44.73\pm0.14$ &  LMXB   &    &  \\
192 & ESO 137-G34  & 248.790 & -58.088 & $1.18\pm0.16$ &  AGN   &    &   \\
193 & IGR J16358-4726\tablenotemark{b} & 248.992 & -47.407 & $1.42\pm0.13$ &  HMXB  &    &  88,89  \\
194 & AX J163904-4642  & 249.768 & -46.707 & $3.95\pm0.13$ &  HMXB  &    &  90 \\
195 & 4U 1636-536  & 250.230 & -53.751 & $38.18\pm0.14$ &  LMXB  &    &   \\
196 & IGR J16418-4532  & 250.465 & -45.534 & $3.50\pm0.14$ &  HMXB  &    &  91,37 \\
197 & GX 340+0  & 251.449 & -45.616 & $28.85\pm0.14$ &  LMXB   &    &   \\
198 & IGR J16465-4507\tablenotemark{b} & 251.648 & -45.118 & $1.66\pm0.14$ &  HMXB   &    &  11,93 \\
199 & IGR J16479-4514  & 252.015 & -45.207 & $3.41\pm0.14$ &  HMXB  &    &  92,93 \\
200 & IGR J16482-3036  & 252.058 & -30.591 & $1.82\pm0.17$ &  AGN   &  2MASX J16481523-3035037  &  44,4 \\
201 & PSR J1649-4349  & 252.373 & -43.823 & $2.24\pm0.15$ &  PSR   &    &   \\
202 & IGR J16500-3307  & 252.493 & -33.113 & $1.12\pm0.16$ &    &  1RXS J164955.1-330713  &  44 \\
203 & NGC 6221  & 253.120 & -59.215 & $1.33\pm0.19$\tablenotemark{e} &  AGN   &    &   \\
204 & NGC 6240  & 253.305 & 2.429 & $3.26\pm0.97$ &  AGN   &    &   \\
205 & MKN 501  & 253.464 & 39.751 & $3.16\pm0.30$ &  AGN   &    &   \\
206 & GRO J1655-40  & 253.499 & -39.844 & $2.85\pm0.14$ &  LMXB  &    &   \\
207 & IGR J16558-5203  & 254.032 & -52.078 & $2.05\pm0.15$ &  AGN   &    &  86,8 \\
208 & IGR J16562-3301  & 254.073 & -33.045 & $1.38\pm0.14$ &    &  SWIFT J1656.3-3302  &   \\
209 & Her X-1  & 254.455 & 35.343 & $89.19\pm0.28$ &  LMXB   &    &   \\
210 & AX J1700.2-4220  & 255.082 & -42.335 & $1.17\pm0.14$ &  HMXB   &    &  63 \\
211 & OAO 1657-415  & 255.199 & -41.656 & $63.72\pm0.14$ &  HMXB   &    &   \\
212 & XTE J1701-462  & 255.232 & -46.197 & $39.26\pm2.50$\tablenotemark{R407} &  LMXB   &    &   \\
213 & GX 339-4  & 255.705 & -48.792 & $46.69\pm0.15$ &  LMXB  &    &   \\
214 & 4U 1700-377  & 255.984 & -37.842 & $193.92\pm0.14$ &  HMXB   &    &   \\
215 & GX 349+2  & 256.431 & -36.421 & $40.74\pm0.13$ &  LMXB   &    &   \\
216 & 4U 1702-429  & 256.566 & -43.037 & $14.93\pm0.15$ &  LMXB   &    &   \\
217 & 4U 1705-32  & 257.223 & -32.322 & $1.93\pm0.12$ &  LMXB   &    &   \\
218 & 4U 1705-440  & 257.234 & -44.102 & $25.00\pm0.14$ &  LMXB   &    &   \\
219 & 1RXS J170849.0-400910  & 257.214 & -40.142 & $1.31\pm0.14$ &  AXP   &    &  \\
220 & IGR J17091-3624  & 257.308 & -36.408 & $5.35\pm0.13$ &  LMXB   &    &  94,24 \\
221 & XTE J1709-267  & 257.386 & -26.658 & $13.92\pm0.67$\tablenotemark{R171} &  LMXB   &    &   \\
222 & XTE J1710-281  & 257.549 & -28.128 & $2.39\pm0.12$ &  LMXB   &    &   \\
223 & RX J1713.7-3946  & 257.991 & -39.862 & $0.61\pm0.14$ &  SNR   &  G347.3-0.5  &   \\
224 & Oph cluster  & 258.114 & -23.347 & $3.66\pm0.13$ &  Cluster   &    &   \\
225 & SAX J1712.6-3739  & 258.153 & -37.645 & $3.97\pm0.13$ &  LMXB   &    &   \\
226 & 4U 1708-40  & 258.120 & -40.858 & $1.02\pm0.14$ &  LMXB   &    &   \\
227 & V2400 Oph  & 258.149 & -24.244 & $2.64\pm0.13$ &  CV   &    &   \\
228 & KS 1716-389  & 259.003 & -38.879 & $5.08\pm1.23$ &  HMXB   &  XTE J1716-389  &  48,49 \\
229 & NGC 6300  & 259.244 & -62.830 & $3.29\pm0.31$ &  AGN   &    &  \\
230 & IGR J17195-4100  & 259.911 & -41.023 & $1.92\pm0.14$ &  CV   &  1RXS J171935.6-410054  &  86,36 \\
231 & XTE J1720-318  & 259.993 & -31.753 & $2.59\pm0.11$ &  LMXB  &    &   \\
232 & IGR J17200-3116  & 260.022 & -31.294 & $1.86\pm0.11$ &  HMXB?   &  1RXS J172006.1-311702  &  86,8 \\
233 & IGR J17204-3554  & 260.087 & -35.900 & $0.79\pm0.12$ &  AGN   &    &  44,27 \\
234 & EXO 1722-363  & 261.295 & -36.282 & $8.18\pm0.12$ &  HMXB   &    &  \\
235 & IGR J17254-3257  & 261.354 & -32.953 & $1.57\pm0.11$ &  LMXB   &  1RXS J172525.5-325717  &  86,70 \\
236 & IGR J17269-4737  & 261.681 & -47.647 & $12.43\pm1.59$\tablenotemark{R364} &  XRB   &  XTE J1726-476  &  73  \\
237 & 4U 1724-30  & 261.888 & -30.804 & $17.09\pm0.10$ &  LMXB  &  Terzan 2  &   \\
238 & IGR J17285-2922  & 262.163 & -29.370 & $3.68\pm0.57$\tablenotemark{R120} &  LMXB?   &  XTE J1728-295  &  86,35 \\
239 & IGR J17303-0601  & 262.579 & -5.971 & $3.54\pm0.30$ &  CV  &  1RXS J173021.5-055933  &  86,72 \\
240 & GX 9+9  & 262.934 & -16.952 & $11.08\pm0.16$ &  LMXB  &    &   \\
241 & GX 354-0  & 262.988 & -33.833 & $35.54\pm0.10$ &  LMXB   &    &   \\
242 & GX 1+4  & 263.011 & -24.747 & $54.39\pm0.11$ &  LMXB   &    &   \\
243 & IGR J17320-1914  & 263.001 & -19.195 & $1.14\pm0.14$ &  Nova   &  V2487 Oph  &  36 \\
244 & IGR J17331-2406  & 263.291 & -24.142 & $1.24\pm0.10$ &     &    &  64 \\
245 & RapidBurster  & 263.349 & -33.387 & $3.73\pm0.11$ &  LMXB   &    &   \\
246 & IGR J17350-2045\tablenotemark{c}  & 263.740 & -20.754 & $0.90\pm0.12$ &    &    &   \\
247 & IGR J17353-3539\tablenotemark{c}  & 263.830 & -35.663 & $0.80\pm0.10$ &    &    &   \\
248 & IGR J17353-3257  & 263.848 & -32.934 & $1.38\pm0.10$ &   &  IGR J17354-3255  &  22 \\
249 & IGR J17364-2711  & 264.117 & -27.199 & $1.60\pm0.30$\tablenotemark{d} &    &    &  82 \\
250 & GRS 1734-292  & 264.371 & -29.139 & $5.18\pm0.10$ &  AGN   &    &  29 \\
251 & IGR J17379-3747\tablenotemark{c}  & 264.465 & -37.774 & $6.34\pm0.91$\tablenotemark{R165} &    &   &  \\
252 & SLX 1735-269  & 264.571 & -26.991 & $10.00\pm0.10$ &  LMXB   &    &   \\
253 & 4U 1735-444  & 264.748 & -44.453 & $25.07\pm0.17$ &  LMXB   &    &  \\
254 & IGR J17391-3021  & 264.812 & -30.355 & $1.08\pm0.09$ &  HMXB  &  XTE J1739-302  &  6,15 \\
255 & GRS 1736-297  & 264.899 & -29.736 & $4.99\pm0.52$\tablenotemark{R409} &     &    &   \\
256 & XTE J1739-285\tablenotemark{b} & 264.975 & -28.496 & $2.01\pm0.10$ &  LMXB   &    &  5 \\
257 & IGR J17402-3656  & 265.087 & -36.936 & $0.87\pm0.12$ &  Open star cluster    &  NGC 6400  &   \\
258 & SLX 1737-282\tablenotemark{b} & 265.168 & -28.313 & $3.66\pm0.09$ &  LMXB  &    &   \\
259 & IGR J17407-2808\tablenotemark{b} & 265.175 & -28.133 & $1.47\pm0.11$ &  HMXB  &  2RXP J174040.9-280852   &  16,10  \\
260 & IGR J17419-2802\tablenotemark{b} & 265.485 & -28.031 & $8.11\pm0.10$\tablenotemark{R409} &     &    &  61  \\
261 & 2E 1739.1-1210   & 265.484 & -12.188 & $1.78\pm0.20$ &  AGN  &  IGR J17418-1212  &   \\
262 & XTE J1743-363  & 265.753 & -36.377 & $2.86\pm0.11$ &  HMXB?   &    &  10 \\
263 & 1E 1740.7-294  & 265.976 & -29.748 & $27.91\pm0.09$ &  LMXB   &    &   \\
264 & IGR J17445-2747  & 266.082 & -27.772 & $4.39\pm0.67$\tablenotemark{R165} &     &    &  44  \\
265 & IGR J17448-3231\tablenotemark{b} \tablenotemark{c} & 266.190 & -32.528 & $0.56\pm0.10$ &    &    &   \\
266 & KS 1741-293\tablenotemark{b} & 266.242 & -29.337 & $5.11\pm0.09$ &  ~~~~~LMXB~~~~~   &    &   \\
267 & GRS 1741.9-2853\tablenotemark{b} & 266.250 & -28.917 & $3.05\pm0.09$ &  LMXB  &    &   \\
268 & IGR J17456-2901\tablenotemark{b} & 266.401 & -29.026 & $5.61\pm0.09$ &   &  Nuclear stellar cluster &  79,76 \\
269 & 1E 1742.8-2853\tablenotemark{b} & 266.500 & -28.914 & $5.95\pm0.09$ &  LMXB?    &    &   \\
270 & A 1742-294  & 266.517 & -29.508 & $11.82\pm0.09$ &  LMXB   &    &  \\
271 & IGR J17464-3213  & 266.564 & -32.237 & $32.30\pm0.10$ &  LMXB  &  H1743-322/XTE J1746-322  &  95 \\
272 & 1E 1743.1-2843\tablenotemark{b} & 266.580 & -28.735 & $5.45\pm0.09$ &  LMXB   &    &   \\
273 & SAX J1747.0-2853\tablenotemark{b} & 266.761 & -28.883 & $3.03\pm0.09$ &  LMXB  &    &  \\
274 & SLX 1744-299/300\tablenotemark{b} & 266.834 & -30.010 & $7.57\pm0.09$ &  LMXB  &    &   \\
275 & IGR J17473-2721  & 266.841 & -27.352 & $5.14\pm0.73$\tablenotemark{R304} &     &    &  74 \\
276 & IGR J17475-2253\tablenotemark{c}  & 266.901 & -22.862 & $0.97\pm0.10$ &    &    &   \\
277 & IGR J17475-2822  & 266.864 & -28.364 & $2.51\pm0.10$ &  Molecular cloud  &  SgrB2  &  21 \\
278 & GX 3+1  & 266.983 & -26.563 & $10.14\pm0.10$ &  LMXB  &    &   \\
279 & A 1744-361  & 267.052 & -36.133 & $16.22\pm0.11$\tablenotemark{R181} &  LMXB   &    &   \\
280 & 4U 1745-203  & 267.217 & -20.359 & $12.03\pm0.54$\tablenotemark{R120} &  LMXB   &    &   \\
281 & AX J1749.1-2733\tablenotemark{b} & 267.275 & -27.550 & $1.48\pm0.10$ &  XRB?    &    &  10  \\
282 & IGR J17488-3253  & 267.223 & -32.907 & $1.34\pm0.10$ &  AGN   &    &  86,8 \\
283 & AX J1749.2-2725\tablenotemark{b} & 267.292 & -27.421 & $1.60\pm0.09$ &  HMXB  &    &   \\
284 & SLX 1746-331  & 267.477 & -33.201 & $0.82\pm0.10$ &  LMXB  &    &   \\
285 & 4U 1746-37  & 267.548 & -37.046 & $2.95\pm0.12$ &  LMXB  &    &   \\
286 & IGR J17505-2644\tablenotemark{c}  & 267.636 & -26.744 & $0.66\pm0.10$ &    &    &   \\
287 & GRS 1747-313  & 267.689 & -31.284 & $1.39\pm0.09$ &  LMXB   &  Terzan 6 &   \\
288 & XTE J1751-305    & 267.816 & -30.616 & $5.91\pm0.61$\tablenotemark{R299} &  LMXB  &    &   \\
289 & IGR J17513-2011  & 267.820 & -20.184 & $1.62\pm0.12$ &  AGN  &    &  44,8  \\
290 & SWIFT J1753.5-0127  & 268.361 & -1.452 & $3.44\pm0.24$ &  LMXB?   &    &  80,81 \\
291 & AX J1754.2-2754  & 268.495 & -28.026 & $2.05\pm0.53$ &  LMXB &    &   \\
292 & IGR J17544-2619  & 268.619 & -26.325 & $0.65\pm0.09$ &  HMXB  &    &  54,31 \\
293 & IGR J17585-3057\tablenotemark{c}  & 269.636 & -30.956 & $0.79\pm0.09$ &    &    &   \\
294 & IGR J17597-2201  & 269.946 & -22.026 & $5.61\pm0.11$ &  LMXB?  &  XTE J1759-220  &  96,97 \\
295 & GX 5-1  & 270.283 & -25.075 & $45.54\pm0.10$ &  LMXB   &    &   \\
296 & GRS 1758-258  & 270.302 & -25.743 & $54.32\pm0.10$ &  LMXB   &    &   \\
297 & GX 9+1  & 270.385 & -20.527 & $15.47\pm0.12$ &  LMXB   &    &   \\
298 & IGR J18027-2016  & 270.666 & -20.283 & $4.07\pm0.12$ &  HMXB  &  IGR/SAX J18027-2017  &  98,99 \\
299 & IGR J18027-1455  & 270.692 & -14.910 & $2.05\pm0.15$ &  AGN  &  1RXS J180245.2-145432(f)  &  86,3 \\
300 & IGR J18048-1455  & 271.180 & -14.925 & $1.06\pm0.15$ &  HMXB   &    &  44,60 \\
301 & XTE J1807-294  & 271.770 & -29.430 & $0.95\pm0.10$ &  LMXB   &    &   \\
302 & SGR 1806-20  & 272.162 & -20.404 & $3.26\pm0.12$ &  SGR    &    &   \\
303 & PSR J1811-1925  & 272.862 & -19.423 & $0.99\pm0.13$ &  PSR/PWN   &  SNR G11.2-0.3  &   \\
304 & IGR J18135-1751  & 273.397 & -17.858 & $1.28\pm0.14$ &  PWN/SNR?   &  HESS J1813-178  &  46 \\
305 & GX 13+1  & 273.629 & -17.155 & $11.96\pm0.14$ &  LMXB   &    &   \\
306 & M 1812-12  & 273.780 & -12.094 & $26.22\pm0.16$ &  LMXB   &    &   \\
307 & GX 17+2  & 274.006 & -14.035 & $55.95\pm0.15$ &  LMXB    &    &  \\
308 & IGR J18170-2511\tablenotemark{c}  & 274.295 & -25.142 & $0.82\pm0.11$ &    &    &   \\
309 & XTE J1817-330  & 274.431 & -33.020 & $7.57\pm0.11$ &  LMXB   &    &   \\
310 & SAX J1818.6-1703  & 274.699 & -17.033 & $1.48\pm0.15$ &  HMXB   &    &  33 \\
311 & AX J1820.5-1434  & 275.112 & -14.564 & $2.19\pm0.16$ &  HMXB   &    &   \\
312 & IGR J18214-1318  & 275.340 & -13.299 & $1.36\pm0.15$ &    &    &  44 \\
313 & 4U 1820-303  & 275.921 & -30.362 & $29.26\pm0.11$ &  LMXB    &    &   \\
314 & IGR J18249-3243  & 276.206 & -32.738 & $0.92\pm0.11$ & AGN  &  PKS 1821-327?   &  9 \\
315 & 4U 1822-000  & 276.312 & 0.007 & $1.59\pm0.16$ &  LMXB   &    &   \\
316 & 4U 1822-371  & 276.447 & -37.106 & $25.61\pm0.14$ &  LMXB   &    &   \\
317 & IGR J18257-0707  & 276.480 & -7.145 & $1.03\pm0.15$ &    &    &  44 \\
318 & LS 5039  & 276.554 & -14.861 & $0.57\pm0.16$ &  HMXB   &    &   \\
319 & GS 1826-24  & 277.367 & -23.798 & $73.65\pm0.13$ &  LMXB   &    &  \\
320 & AX J183039-1002  & 277.660 & -10.049 & $0.84\pm0.16$ &    &    &   \\
321 & IGR J18325-0756  & 278.112 & -7.938 & $1.89\pm0.15$ &     &    &  100 \\
322 & SNR 021.5-00.9  & 278.394 & -10.572 & $2.88\pm0.16$ &  SNR   &    &   \\
323 & PKS 1830-211  & 278.421 & -21.068 & $2.34\pm0.15$ &  AGN   &    &   \\
324 & 3C382  & 278.738 & 32.660 & $2.77\pm1.22$ &  AGN   &    &   \\
325 & RX J1832-33  & 278.933 & -32.990 & $9.59\pm0.13$ &  LMXB  &    &   \\
326 & AX J1838.0-0655  & 279.503 & -6.911 & $2.03\pm0.15$ &  SNR/PWN   &  HESS J1837-069  &  47 \\
327 & ESO 103-G035  & 279.632 & -65.422 & $4.64\pm0.94$ &  AGN   &    &   \\
328 & Ser X-1  & 279.991 & 5.041 & $10.27\pm0.13$ &  LMXB  &    &   \\
329 & IGR J18410-0535  & 280.262 & -5.577 & $1.11\pm0.15$ &  HMXB   &  AX J1841.0-0536  &  19 \\
330 & 1E 1841-045  & 280.329 & -4.938 & $2.48\pm0.15$ &  PSR/PWN  &    &   \\
331 & 3C390.3  & 280.578 & 79.763 & $4.30\pm0.44$ &  AGN   &    &   \\
332 & AX J1845.0-0433  & 281.253 & -4.574 & $1.43\pm0.14$ &  HMXB       &    &  40 \\
333 & GS 1843+00  & 281.404 & 0.868 & $4.63\pm0.13$ &  HMXB   &    &   \\
334 & PSR J1846-0258  & 281.613 & -2.983 & $1.57\pm0.15$ &  PSR/PWN  &  AXP?  &  \\
335 & A 1845-024  & 282.048 & -2.426 & $0.93\pm0.13$ &  HMXB   &    &   \\
336 & IGR J18483-0311  & 282.071 & -3.172 & $3.18\pm0.14$ &    &    &  71 \\
337 & IGR J18486-0047\tablenotemark{c}  & 282.104 & -0.787 & $1.07\pm0.13$ &    &    &   \\
338 & IGR J18490-0000  & 282.258 & -0.013 & $1.13\pm0.13$ &    &    &  38 \\
339 & 4U 1850-087  & 283.265 & -8.702 & $5.02\pm0.15$ &  ~~~~~LMXB~~~~~  &    &   \\
340 & IGR J18539+0727  & 283.500 & 7.488 & $0.69\pm0.12$ &  LMXB?  &     &  30,24 \\
341 & 4U 1849-31  & 283.761 & -31.155 & $6.40\pm0.18$ &  CV  &  V1223 Sgr  &   \\
342 & XTE J1855-026  & 283.870 & -2.601 & $10.61\pm0.13$ &  HMXB   &    &  \\
343 & IGR J18559+1535  & 283.987 & 15.629 & $1.59\pm0.16$ &  AGN   &  2E 1853.7+1534  &  32,8 \\
344 & IGR J18578-3405  & 284.469 & -34.096 & $4.01\pm0.64$\tablenotemark{R408} &  AGN?   &   &  \\
345 & XTE J1858+034  & 284.673 & 3.437 & $12.09\pm0.12$ &  HMXB   &    &   \\
346 & HETE J19001-2455  & 285.039 & -24.917 & $7.03\pm0.21$ &  LMXB   &    &   \\
347 & XTE J1901+014  & 285.415 & 1.447 & $2.59\pm0.12$ &  HMXB?  &    &  69  \\
348 & 4U 1901+03  & 285.917 & 3.207 & $31.22\pm0.12$ &  HMXB   &    &   \\
349 & SGR 1900+14  & 286.839 & 9.322 & $0.91\pm0.11$ &  SGR   &    &   \\
350 & XTE J1908+094  & 287.219 & 9.374 & $1.45\pm0.12$ &  LMXB   &    &   \\
351 & 4U 1907+097  & 287.406 & 9.833 & $14.59\pm0.12$ &  HMXB   &    &   \\
352 & IGR J19108+0917  & 287.641 & 9.312 & $2.70\pm0.55$ &    &    &   \\
353 & X 1908+075  & 287.701 & 7.595 & $13.04\pm0.12$ &  HMXB   &    &   \\
354 & Aql X-1  & 287.814 & 0.584 & $12.30\pm0.12$ &  LMXB   &    &   \\
355 & SS 433  & 287.957 & 4.979 & $8.78\pm0.11$ &  HMXB   &    &  \\
356 & IGR J19140+098  & 288.526 & 9.885 & $8.99\pm0.12$ &  HMXB  &  IGR J19140+0951  &  53,42  \\
357 & GRS 1915+105  & 288.801 & 10.947 & $261.49\pm0.12$ &  LMXB   &    &   \\
358 & 4U 1916-053  & 289.686 & -5.247 & $8.04\pm0.16$ &  LMXB   &    &   \\
359 & SWIFT J1922.7-1716  & 290.615 & -17.300 & $8.38\pm1.23$\tablenotemark{R309} &     &    &  65 \\
360 & 1H 1934-063  & 294.422 & -6.240 & $1.24\pm0.22$ &  AGN   &    &   \\
361 & RX J1940.2-1025  & 295.050 & -10.446 & $2.32\pm0.28$ &  CV   &  V1432 Aql  &  36 \\
362 & NGC 6814  & 295.685 & -10.331 & $3.30\pm0.29$ &  AGN   &    &   \\
363 & XSS J19459+4508  & 296.887 & 44.883 & $1.06\pm0.31$ &  AGN  &  IGR J19473+4452 &  23,2  \\
 & & & & & & 2MASX J19471938+4449425 & \\
364 & KS 1947+300  & 297.397 & 30.211 & $13.11\pm0.30$ &  HMXB   &    &   \\
365 & 3C403  & 298.024 & 2.445 & $5.18\pm1.56$ &  AGN   &    &   \\
366 & 4U 1954+319  & 298.933 & 32.094 & $4.76\pm0.27$ &  HMXB   &    &   \\
367 & Cyg X-1  & 299.588 & 35.202 & $736.49\pm0.24$ &  HMXB   &    &   \\
368 & Cygnus A  & 299.863 & 40.736 & $4.03\pm0.23$ &  AGN   &    &   \\
369 & SWIFT J2000.6+3210  & 300.101 & 32.166 & $1.89\pm0.26$ &  HMXB   &  IGR J20006+3210    &  65,66  \\
370 & IGR J20187+4041  & 304.647 & 40.706 & $1.32\pm0.19$ &  AGN   &  2MASX J20183871+4041003 &   26,39 \\
 & & & & & &  IGR J2018+4043 & \\
371 & IGR J20286+2544  & 307.140 & 25.746 & $2.31\pm0.41$ &  AGN   &  MCG +04-48-002  &  9,7 \\
372 & EXO 2030+375  & 308.062 & 37.638 & $33.45\pm0.19$ &  HMXB   &    &   \\
373 & Cyg X-3  & 308.108 & 40.959 & $160.81\pm0.18$ &  HMXB   &    &   \\
374 & 4C +74.26  & 310.576 & 75.141 & $2.74\pm0.85$ &  AGN   &    &   \\
375 & MRK 509  & 311.036 & -10.727 & $3.85\pm0.58$ &  AGN    &    &   \\
376 & IGR J20569-4940  & 314.215 & 49.679 & $1.03\pm0.22$ &    &  3A 2056+493  &   \\
377 & SAX J2103.5+4545  & 315.901 & 45.753 & $10.47\pm0.18$ &  HMXB   &    &  \\
378 & S5 2116+81  & 318.736 & 82.045 & $2.48\pm0.67$ &  AGN   &    &   \\
379 & IGR J21178+5139  & 319.441 & 51.650 & $1.35\pm0.24$ &  AGN   &  2MASX J21175311+5139034  &  9 \\
380 & IGR J21237+4218  & 320.960 & 42.316 & $1.03\pm0.18$ &  CV   &  V2069 Cyg  &   \\
381 & IGR J21247+5058  & 321.161 & 50.973 & $5.98\pm0.24$ &  AGN   &    &  86,3 \\
382 & IGR J21277+5656  & 321.930 & 56.943 & $1.87\pm0.31$ &  AGN   &    &  32 \\
383 & 4U 2127+119  & 322.502 & 12.172 & $3.72\pm0.47$ &  LMXB  &    &   \\
384 & CV Cyg  & 323.449 & 51.122 & $3.15\pm0.26$ &  CV   &    &  36 \\
385 & IGR J21343+4738  & 323.625 & 47.614 & $1.09\pm0.21$ &    &  1RXS J213419.6+473810  &   \\
386 & SS Cyg  & 325.711 & 43.574 & $2.89\pm0.22$ &  CV  &    &   \\
387 & Cyg X-2  & 326.170 & 38.319 & $25.81\pm0.23$ &  ~~~~~LMXB~~~~~   &    &   \\
388 & NGC 7172  & 330.490 & -31.864 & $4.18\pm0.35$ &  AGN   &    &   \\
389 & BL Lac  & 330.645 & 42.267 & $1.31\pm0.26$ &  AGN   &    &  1 \\
390 & 4U 2206+543  & 331.992 & 54.513 & $6.40\pm0.25$ &  HMXB   &    &   \\
391 & FO Aqr  & 334.402 & -8.301 & $3.28\pm0.99$\tablenotemark{R25} &  CV   &    &   \\
392 & NGC 7314  & 338.890 & -26.021 & $1.51\pm0.40$ &  AGN   &    &   \\
393 & IGR J22367-1231  & 339.176 & -12.539 & $1.83\pm0.39$ &  AGN   &  MRK 915  &   \\
394 & IGR J22517+2218\tablenotemark{c}  & 342.939 & 22.316 & $2.45\pm0.43$\tablenotemark{R316} &    &    &    \\
395 & 3C 454.3  & 343.490 & 16.143 & $9.73\pm0.38$ &  AGN   &    &  1 \\
396 & MR 2251-178  & 343.465 & -17.616 & $3.32\pm0.34$ &  AGN    &    &   \\
397 & NGC 7469  & 345.825 & 8.879 & $3.31\pm0.54$ &  AGN   &    &   \\
398 & MRK 926  & 346.166 & -8.689 & $2.49\pm0.37$ &  AGN  &    &   \\
399 & Cas A  & 350.846 & 58.813 & $3.68\pm0.11$ &  SNR   &    &   \\
400 & IGR J23523+5844\tablenotemark{c}  & 358.079 & 58.745 & $0.62\pm0.10$ &    &    &   \\
\enddata 
\tablenotetext{a}{Thermal emission dominates.}
\tablenotetext{b}{Spatial confusion, source flux should be taken with the caution.}
\tablenotetext{c}{Newly discovered sources in this survey.}
\tablenotetext{d}{Flux measured in the energy range $15-25$~keV over 10 s during the burst \citep{cheletal06}.}
\tablenotetext{e}{May contain flux from ESO 138-G1 (AGN, $\sim6.5$\arcmin from NGC6221)} 
\tablenotetext{R}{Source flux was measured on the map averaged over the indicated spacecraft revolution.}
\end{deluxetable}

\end{document}